\documentclass[twocolumn]{autart}    

\usepackage{graphicx}          

\usepackage{amsfonts}
\usepackage{amsmath}
\usepackage{cite}
\usepackage{color}
\usepackage{enumitem}

\newtheorem{theorem}{Theorem}
\newtheorem{lemma}{Lemma}

\newtheorem{remark}{Remark}
\newtheorem{proposition}{Proposition}

\newtheorem{example}{Example}

\newcommand{\R}{\mathbb{R}}

\newcommand{\T}{\text{T}}

\usepackage{xcolor}

 \usepackage{mathptmx}

 \allowdisplaybreaks
\begin{document}

\begin{frontmatter}

\title{Adaptive Override Control under High-Relative-Degree Nonovershooting Constraints} 

\thanks[footnoteinfo]{This paper was not presented at any IFAC
meeting.}

\author[NTU,TJU]{Ziliang Lyu}\ead{ziliang\_lyu@outlook.com},
\author[UCSD]{Miroslav Krstic}\ead{mkrstic@ucsd.edu},
\author[NUS]{Kaixin Lu}\ead{lukaixin@nus.edu.sg},
\author[TJU]{Yiguang Hong}\ead{yghong@iss.ac.cn},
\author[NTU]{Lihua Xie}\ead{elhxie@ntu.edu.sg}

\address[NTU]{School of Electrical and Electronic Engineering, Nanyang Technological University, Singapore 639798, Singapore}
\address[TJU]{Department of Control Science and Engineering, Tongji University, Shanghai 200092, China}
\address[UCSD]{Department of Mechanical and Aerospace Engineering, University of California at San Diego, La Jolla, CA, 92093-0411 USA}
\address[NUS]{Department of Biomedical Engineering, National University of Singapore, Singapore}


\begin{abstract}                          
This paper considers the problem of adaptively overriding unsafe actions of a nominal controller in the presence of high-relative-degree nonovershooting constraints and parametric uncertainties. To prevent the design from being coupled with high-order derivatives of the parameter estimation error, we adopt a modular design approach in which the controller and the parameter identifier are designed separately. The controller module ensures that any safety violations caused by parametric uncertainties remain bounded, provided that the parameter estimation error and its first-order derivative are either bounded or square-integrable. The identifier module, in turn, guarantees that these requirements on the parameter estimation error are satisfied. Both theoretical analysis and simulation results demonstrate that the closed-loop safety violation is bounded by a tunable function of the initial estimation error. Moreover, as time increases, the parameter estimate converges to the true value, and the amount of safety violation decreases accordingly.
\end{abstract}

\end{frontmatter}

\section{Introduction}

In flight, automotive, or manufacturing systems, overshoot can cause serious damage to the plant or its environment. Traditional nonovershooting control problems focus on forcing the system output to track a nominal reference trajectory (or setpoint) while preventing the output from exceeding the reference; see \cite{phillips1988conditions, el1993influence, deodhare1990design, bement2004construction} for linear systems, and \cite{krstic2006nonovershoot, li2020mean, li2023prescribed} for nonlinear systems. In these results, the reference trajectory is also the boundary of the time-varying safe set, and thus, one can design a controller to achieve the nominal control objective while avoiding safety violations. In 2006, a nonovershooting backstepping method was proposed in \cite{krstic2006nonovershoot} for the first time for strict-feedback nonlinear systems such that the closed-loop system satisfies
\begin{flalign}
    &\dot{h}_i=-c_ih_i+h_{i+1},\;\;i=1,\ldots,n-1,
        \label{eq:nonovershoot-backstepping-A}\\
    &\dot{h}_n=-c_nh_n,
        \label{eq:nonovershoot-backstepping-B}
\end{flalign}
where $h_1(x,t)$ is a sufficiently smooth function that captures the relationship between the output and the safety constraint, and $c_1$, \ldots, $c_n$ are positive constants. By taking $h(x,t)=h_1(x,t)$, the hyperplane described by (\ref{eq:nonovershoot-backstepping-A})-(\ref{eq:nonovershoot-backstepping-B}) is actually the boundary of the invariant subspace described by
\begin{flalign}
    h^{(n)}+a_{1}h^{(n-1)}+\cdots+a_{n-1}h^{(1)}+a_nh\geq0,
\end{flalign}
($a_i$'s are positive constants related to $c_i$'s) which is now known as the high-relative-degree barrier function \cite{ames2019control, wu2015safety, nguyen2016exponential, xu2018constrained, tan2021high, xiao2021high, lyu2023small, abel2023prescribed}---one of the most important concepts in the field of safety-critical control.

Unlike the traditional nonovershooting control problems studied in \cite{phillips1988conditions, el1993influence, deodhare1990design, bement2004construction, krstic2006nonovershoot}, achieving nominal control objectives in general safety-critical control problems may lead to safety violations. A representative example is the adaptive cruise control problem \cite{ ames2016control }, where two vehicles are already at the minimum safe distance, yet the following car is commanded to accelerate beyond the speed of the leading vehicle. Therefore, it is necessary to introduce a safety filter to modify the nominal control input when safety constraints are at risk of being violated. Typically, a safety filter consists of two parts: 1) an override controller, which serves as a ``baseline'' that characterizes what kind of control input is safe, and 2) an override mechanism, which determines when to activate the override controller to modify the nominal controller's unsafe actions.

Control barrier functions (CBFs) play an important role in the design of override controllers. The concept of CBFs were first introduced in \cite{wieland2007constructive}, based on the combination of control Lyapunov functions (CLFs) for stabilization \cite{ sontag1989universal} and barrier certificates for safety verification \cite{ prajna2007framework}. This CBF formulation requires all superlevel sets of a CBF to be forward invariant and is overly conservative. Later, \cite{ames2014control, ames2016control, xu2015robustness} refined the CBF inequality to allow that only the zero superlevel set is forward invariant. Essentially, CBF methods synthesize a safety-critical controller using the relationship between the control input and the CBF's value and change. This gives rise to the high-relative-degree problem, in which the influence of the control input on the CBF's value and change becomes explicit only after differentiating the CBF multiple times. To handle safety constraints with arbitrary relative degree, \cite{ nguyen2016exponential } introduced a high-relative-degree CBF concept. As noted earlier, a similar idea appeared in the earlier work of \cite{krstic2006nonovershoot}; however, the key distinction lies in the formulation: \cite{nguyen2016exponential} employed a ``modern CBF language'' to explicitly define a convex set of control input safer than the override controller.

The CBF quadratic program (CBF-QP) \cite{ ames2014control} is an override mechanism that maximizes liveness at each time and state. In plain language, CBF-QP safety filters assess whether the nominal controller is at the risk of safety violations, based on whether it is less safe than the override controller, rather than using the earlier criterion (e.g., \cite{wieland2007constructive}) of checking whether the system state enters a predefined neighborhood of the unsafe region. When such a risk is detected, the override controller is activated to minimally modify the nominal control input. This QP-based override strategy offers two key advantages. First, the resulting closed-loop controller is locally Lipschiz continuous, provided that the QP is feasible \cite{xu2015robustness}. Second, the closed-loop controller is pointwise optimal \cite{ ames2016control }, and can be further enhanced to be infinite-horizon optimal by increasing the gain of the override part  twice or more \cite{krstic2023inverse}.

Real-world control systems often involve unknown parameters. Adaptive control has proven highly effective for handling such parametric uncertainties with online-acquired information about the unknown parameters \cite{krstic1995nonlinear}. In the context of safety-critical control, \cite{taylor2020adaptive} introduced an adaptive control barrier function (ACBF) for the first time, with an idea of combining adaptive control Lyapunov function \cite{krstic1995aclf} and barrier certificates \cite{prajna2007framework}. Another ACBF with less conservativeness was later proposed in \cite{lopez2020robust}, where only the zero superlevel set of the ACBF is required to be forward invariant. Based on ACBFs, further adaptive safety-critical control results have been developed in \cite{isaly2021adaptive, wang2024adaptive}. However, all these results are restricted to relative-degree-one safety constraints. Although safety-critical control under high-relative-degree constraints has been widely studied for nearly two decades (initially in the context of nonovershooting constraints \cite{krstic2006nonovershoot} and later for more general safety constraints \cite{nguyen2016exponential, xu2018constrained, tan2021high, xiao2021high, lyu2023small, abel2023prescribed, kim2025constant, cohen2024safety, taylor2022safe, ong2024rectified, hsu2015control, wu2015safety}), the adaptive control under this setting remains unsolved. A recent attempt was made in \cite{cohen2023modular} for systems satisfying the matching condition, i.e., the unknown parameters enter the chain of CBF inequalities at the same place as the control input. Under such assumptions, the high-relative-degree problem can be reduced to a relative-degree-one problem studied in \cite{taylor2020adaptive, lopez2020robust}.

For strict-feedback systems, nonovershooting constraints belong to a special case of high-relative-degree safety constraints. This paper focuses on the adaptive nonovershooting control of strict-feedback systems with unknown parameters. Our adaptive nonovershooting controller can be directly used as an override controller in the QP-based safety-critical control framework. Compared to \cite{taylor2020adaptive, lopez2020robust }, our result can handle high-relative-degree safety constraints. Compared to \cite{cohen2023modular}, our parametric uncertainty is not matched. The conventional approach to handling high-relative-degree constraints is to repeatedly differentiate a barrier function candidate until the control input appears explicitly \cite{nguyen2016exponential, xu2018constrained, tan2021high, xiao2021high, lyu2023small}. However, because ACBFs are coupled with the parameter estimation error \cite{taylor2020adaptive, lopez2020robust}, differentiating an ACBF repeatedly along the solutions of a control system with unmatched uncertainties may introduce a series of higher-order derivatives of the parameter estimation error, which poses significant challenges to both controller and adaptation law design.

This paper is an adaptive extension of \cite{krstic2006nonovershoot}. To prevent the design procedure from being coupled with high-order derivatives of the parameter estimation error, the controller and identifier modules are designed separately. For the controller module, we propose a backstepping method to design a controller such that the closed-loop safety violation caused by parametric uncertainties is bounded, provided that the parameter estimation error is bounded and its first-order derivative is either bounded or square-integrable. To make the parameter estimation satisfy these requirements, we develop four parameter identifiers: the passive error-observer scheme, the swapping error-observer scheme, the passive plant-observer scheme, and the swapping plant-observer scheme. Compared to passive schemes, which can only guarantee that the derivative of the parameter estimation error is bounded, swapping schemes can guarantee that the derivative is both bounded and square-integrable. Compared to error-observer schemes, plant-observer schemes are better suited for cases involving multiple barrier functions.

\textbf{Notation:}
A function $\gamma:\R_{\geq0}\rightarrow\R_{\geq0}$ is called a $K$-function if it is continuous, strictly increasing, and satisfies $\gamma(0)=0$. It is further called a $K_\infty$-function if $\gamma(s)\rightarrow\infty$ as $s\rightarrow\infty$. A function $\beta:\R_{\geq0}\times\R_{\geq0}\rightarrow\R_{\geq0}$ is a $KL$-function if, for each fixed $t\geq0$, the mapping $s\mapsto\beta(s,t)$ is a $K$-function, and for each fixed $s\geq0$, the mapping $t\mapsto\beta(s,t)$ decreases to zero as $t\rightarrow\infty$. We use $|x|$ to denote the Euclidean norm of a vector $x$, and $|X|_F$ and $|X|_2$ to denote the Frobenius norm and the induced 2-norm of a matrix $X$, respectively. The spaces $L_\infty[0,t_f)$ and $L_2[0,t_f)$ denote the sets of signals that are globally bounded and square-integrable on $[0,t_f)$, respectively. For a signal $x$ belonging to $L_\infty[0,t_f)$ or $L_2[0,t_f)$, we denote by $\|x\|_\infty$ and $\|x\|_2$ the $L_\infty$ and $L_2$ norm of its truncation on $[0,t_f)$, respectively.

\section{Problem Statement}\label{sec:state-feedback}

Consider the strict-feedback system
\begin{flalign}\label{eq:str-fb-sys}
    &\dot{x}_i=x_{i+1}+\varphi_i(\bar{x}_i)^T\theta,\;\;1\leq i\leq n-1
     \nonumber\\
    &\dot{x}_n=u+\varphi_n(x)^\T\theta
     \nonumber\\
    &y=x_1
\end{flalign}
where $\bar{x}_i=[x_1,\ldots,x_i]^T$, $x=[x_1,\ldots,x_n]^T$, $\theta\in\R^p$ is a vector of unknown constant parameters, and the components of $\varphi_i$, $i=1,\ldots,n$, are smooth functions. Denote by $x(t,x_0)$ the state trajectory of (\ref{eq:str-fb-sys}) with initial state $x_0:=x(0)$.

In this paper, the output $y$ is required to satisfy the safety constraint
\begin{flalign}\label{eq:safety-constraint}
    h_1(x,t):=y(t)-r(t)\geq0,   \;\;\forall t\geq0
\end{flalign}
where $r(t)$ is a function that can be used to capture the boundary of a safe set. For notational simplicity, we write $h_1(x,t)$ as $h_1(t)$ whenever there is no ambiguity from the context. In this paper, we assume that $r(t)$ and its first $n$ derivatives are known and bounded, and, in addition, $r^{(n)}(t)$ is piecewise continuous.

\begin{remark}\em
Constraint \eqref{eq:safety-constraint} is actually a ``nonundershooting'' constraint, which differs from the conventional formulation in the nonovershooting control literature \cite{krstic2006nonovershoot, li2020mean, li2023prescribed}, where the output is required to satisfy $y(t) \leq r(t)$. We adopt the form \eqref{eq:safety-constraint} to align with the standard convention in the CBF literature \cite{ames2019control, ames2016control}, where safety constraint is expressed using an inequality of the form $h_1(x,t)\geq0$. Whenever no confusion arises, we refer to \eqref{eq:safety-constraint} as a nonovershooting constraint.
\end{remark}

Due to the unknown parameter $\theta$, the output $y$ may violate the safety constraint (\ref{eq:safety-constraint}). Let $\hat{\theta}$ be an estimate of $\theta$, and let $\tilde\theta=\theta-\hat{\theta}$ be the parameter estimation error. Our objective is to design an adaptive nonovershooting controller of the form
\begin{flalign}\label{eq:general-form-nonovershoot-control}
    u=\bar{u}(x,r,\hat{\theta}),\;\;
    \dot{\hat\theta}&=\tau(x,r)
\end{flalign}
such that the closed-loop system satisfies the following properties:
\begin{itemize}
  \item[(P1)] The closed-loop system is practically safe in the sense of
\begin{flalign}\label{eq:h1-low-bd-in-formulation}
    h_1(x,t)\geq-\rho(|\tilde{\theta}(0)|),\;\;\forall t\geq0
\end{flalign}
for all $x_0$ satisfying $h_1(0)\geq0$, where $\rho$ is a $K_\infty$-function that can be made arbitrarily small by appropriately tuning design parameters.
  \item[(P2)] All solutions of the closed-loop system are globally uniformly bounded and
\begin{flalign}
    \lim_{t\rightarrow+\infty}h_1(x,t)=0.
\end{flalign}
\end{itemize}

To more intuitively illustrate the importance of properties (P1)-(P2) in adaptive safety-critical control, we introduce the following proposition.

\begin{proposition}\label{prop:delta-epsilon-KL-convergence}
If properties (P1)-(P2) hold, then
\begin{flalign}\label{eq:KL-bd}
    h_1(x,t)\geq-\beta(|\tilde{\theta}(0)|,t),\;\;\forall h_1(0)\geq0,\;\;\forall t\geq0
\end{flalign}
where $\beta$ is a $KL$-function satisfying $\beta(s,0)=(a+1)\rho(s)$ with $a>0$ an arbitrary constant.
\end{proposition}

Proposition \ref{prop:delta-epsilon-KL-convergence} is inspired by \cite[Proposition 2.5]{lin1996smooth} and the proof is given in Appendix \ref{appendix-A}.

\begin{remark}\em
The $KL$ lower bound in (\ref{eq:KL-bd}) implies that, if properties (P1)-(P2) hold, the safety violation bound is a tunable, increasing function of the initial estimation error, and the effect of the parametric uncertainty on the closed-loop safety is reduced as $t$ increases.
\end{remark}

\begin{remark}\em
Note that adaptive nonovershooting controller (\ref{eq:general-form-nonovershoot-control}) can serve as the override controller in a pointwise minimally optimal safety filter. Specifically, let $u_0$ be a controller used to accomplish the nominal control task and $U_{\rm safe}$ be a convex set of safe control inputs whose boundary is characterized by the baseline controller $\bar{u}(x,r,\hat{\theta})$. Then one can use the following pointwise minimally optimal safety filter to override the unsafe actions of $u_0$:
\begin{flalign}\label{general-safety-filter}
    u=\arg\min_{u\in\R^n} |u-u_0|^2,\;\;\text{s.t.}\;\;u\in U_{\rm safe}.
\end{flalign}
Intuitively, (\ref{general-safety-filter}) ``projects'' the unsafe control into the safe control set $U_{\rm safe}$. In safe control, both $U_{\rm safe}$ and $\bar{u}$ can be constructed by CBF approaches \cite{ames2014control, ames2016control, krstic2006nonovershoot}.
\end{remark}

\section{Controller Module Design}\label{eq:ISSf-design}

Inspired by \cite{krstic2006nonovershoot}, this section designs a nonovershooting backstepping controller such that the closed-loop safety violation is bounded by the $L_\infty$ norm of the parameter estimation error $\tilde{\theta}$ and either the $L_2$ or $L_\infty$ norm of its derivative $\dot{\tilde\theta}=-\dot{\hat{\theta}}$.

Introduce a change of coordinates
\begin{flalign}
    h_1&=y-r
    \label{eq:coordinate-change-A}\\
    h_i&=x_i-\alpha_{i-1}-r^{(i-1)},\;\;i=2,\ldots,n
    \label{eq:coordinate-change-B}\\
    h_{n+1}&=u-\bar{u}
\end{flalign}
where $\alpha_{i-1}$'s are virtual controllers and $\bar{u}$ is the override controller. Design
\begin{flalign}
    \alpha_0
    &=0,
    \label{eq:alpha-0}\\
    \alpha_i
    &=-s_ih_i-w_i^\T\hat\theta
     \nonumber\\
    &\;\;\;\;\;\;\;\;
        +\sum_{k=1}^{i-1}\bigg(\frac{\partial\alpha_{i-1}}{\partial x_k}x_{k+1}+\frac{\partial\alpha_{i-1}}{\partial r^{(k-1)}}r^{(k)}\bigg),
    \label{eq:alpha-i}\\
    s_i
    &=c_i+\kappa_i|w_i|^2+g_i\bigg|\frac{\partial\alpha_{i-1}}{\partial\hat\theta}\bigg|^2,
        \label{eq:si}\\
    w_i
    &=\varphi_i-\sum_{j=1}^{i-1}\frac{\partial\alpha_{i-1}}{\partial x_j}\varphi_j,\;\;\;\;i=1,\ldots,n,
        \label{eq:wi}\\
    \bar{u}
    &=\alpha_n+r^{(n)}
    \label{eq:state-fb-controller}
\end{flalign}
where $c_i,\kappa_i,g_i>0$ are design parameters. By (\ref{eq:coordinate-change-A})-(\ref{eq:state-fb-controller}), the plant (\ref{eq:str-fb-sys}) is transformed into the error system
\begin{flalign}\label{eq:dot-zi-final-state-fb}
    &\dot{h}_i=-s_ih_i+h_{i+1}+w_i^\T\tilde\theta-\dfrac{\partial\alpha_{i-1}}{\partial\hat{\theta}}\dot{\hat\theta},\;\;i=1,\ldots,n-1
     \nonumber\\
    &\dot{h}_n=-s_nh_n+u-\bar{u}+w_n^\T\tilde\theta-\dfrac{\partial\alpha_{n-1}}{\partial\hat\theta}\dot{\hat\theta}
\end{flalign}
Let $h=[h_1,\ldots,h_n]^\T$. Rewrite (\ref{eq:dot-zi-final-state-fb}) in a compact form
\begin{flalign}\label{eq:dot-z-A}
    \dot{h}
    &=A(h,\hat\theta,t)h
        +e_n(u-\bar{u})
        \nonumber\\
    &\;\;\;\;\;\;\;\;\;\;\;\;
        +W(h,\hat\theta,t)^\T\tilde\theta
        +Q(h,\hat\theta,t)^\T\dot{\hat\theta}
\end{flalign}
where $e_n=[0,0,\ldots,1]$, and
\begin{flalign}
    A(h,\hat\theta,t)
    &=
    \left[
      \begin{array}{cccc}
        -s_1 & 1    & \cdots & 0 \\
        0    & -s_2 & \ddots & \vdots \\
        \vdots & \ddots & \ddots & 1 \\
        0    & \cdots & 0 & -s_n \\
      \end{array}
    \right]\in\R^{n\times n},
        \label{eq:def-A}\\
    W(h,\hat\theta,t)
    &=
    \left[
      \begin{array}{cccc}
        w_1 & w_2 & \cdots & w_n \\
      \end{array}
    \right]\in\R^{p\times n},
        \label{eq:def-W}\\
    Q(h,\hat\theta,t)
    &=
    \left[
      \begin{array}{cccc}
        0 & -\bigg(\dfrac{\partial\alpha_1}{\partial\hat\theta}\bigg)^\T & \cdots & -\bigg(\dfrac{\partial\alpha_{n-1}}{\partial\hat\theta}\bigg)^\T \\
      \end{array}
    \right]\in\R^{p\times n}.
        \label{eq:def-Q}
\end{flalign}


Let
\begin{flalign}
    \underline{c}_i
    &=-\Big[x_i(0)-\alpha_{i-1}(0)-r^{(i-1)}(0)\Big]^{-1}
        \nonumber\\
    &\;\;\;\;
        \times\Bigg[x_{i+1}(0)
    -r^{(i)}(0)
    +w_i(0)^\T\hat\theta(0)
        \nonumber\\
    &\;\;\;\;\;\;\;
    -\sum_{j=1}^{i-1}\Big(\dfrac{\partial\alpha_{i-1}(0)}{\partial x_j}x_{j+1}(0)
    +\dfrac{\partial\alpha_{i-1}(0)}{\partial r^{(j-1)}}r^{(j)}(0)\Big)\Bigg]
\end{flalign}
for $i=1,\ldots,n$. Take
\begin{equation}\label{eq:underline-c-kappa-g}
    \underline{c}=\min_{i=1,\ldots,n}c_i,\;\;\;\;
    \underline{\kappa}=\min_{i=1,\ldots,n}\kappa_i,\;\;\;\;
    \underline{g}=\min_{i=2,\ldots,n}g_i
\end{equation}
where $c_i$, $\kappa_i$ and $g_i$ are design parameters associated with the nonlinear damping function (\ref{eq:si}).

\begin{lemma}\label{lem:ISSf-state-feedback}
Suppose that the solutions of the closed-loop system (\ref{eq:dot-z-A}) with the nonovershooting override control $u=\bar{u}$ are defined on $[0,t_f)$. Then, by taking design parameters
\begin{equation}\label{eq:choice-of-ci}
c_i\geq\max\{\underline{c}_i,0\}, \; \text{for} \; i = 1, \ldots, n-1,\; \text{and} \;\; c_n > 0,
\end{equation}
the following properties hold for any $y(0)\geq r(0)$:
\begin{itemize}
  \item[(i)] If $\tilde\theta,\dot{\hat\theta}\in L_\infty[0,t_f)$, then $h,x\in L_\infty[0,t_f)$, and
\begin{equation}\label{eq:zi-L_infty-bd}
    -h_1^*\leq h_1(t)\leq \e^{-\underline{c}t}\sum_{i=1}^n\frac{t^{i-1}}{(i-1)!}h_i(0)+h_1^*
\end{equation}
where
\begin{equation}\label{eq:h-star-bd-Linfty}
    h_1^*=\frac{\underline{c}^{n}-1}{\underline{c}^{n-1}(\underline{c}-1)}
    \Bigg(\frac{\|\tilde\theta\|_\infty}{2\sqrt{\underline{c}\,\underline{\kappa}}}+\frac{\|\dot{\hat\theta}\|_\infty}{2\sqrt{\underline{c}\,\underline{g}}}\Bigg).
\end{equation}
  \item[(ii)] If $\tilde\theta\in L_\infty[0,t_f)$ and $\dot{\hat\theta}\in L_2[0,t_f)$, then $h,x \in L_\infty[0,t_f)$, and $h_1$ satisfies (\ref{eq:zi-L_infty-bd}) with
\begin{equation}\label{eq:h-star-bd-L2}
    h_1^*=\frac{\underline{c}^{n}-1}{\underline{c}^{n-1}(\underline{c}-1)}
    \scalebox{1.1}{\Bigg(}\frac{\|\tilde\theta\|_\infty}{2\sqrt{\underline{c}\,\underline{\kappa}}}+\frac{\|\dot{\hat\theta}\|_2}{\sqrt{2\underline{g}}}\scalebox{1.1}{\Bigg)}.
\end{equation}
\end{itemize}
\end{lemma}

\noindent
\textbf{Proof.}
By applying the variation of constants formula to (\ref{eq:dot-zi-final-state-fb}),
\begin{flalign}\label{eq:zi-equal}
    h_i(t)
    &=h_i(0)\e^{-\int_0^ts_i(r)dr}
        \nonumber\\
    &\;\;\;\;\;\;
        +\int_0^t\e^{-\int_\tau^ts_i(r)dr}\bigg[w_i^T\tilde\theta-\frac{\partial\alpha_{i-1}}{\partial\hat\theta}\dot{\hat\theta}\bigg]d\tau
        \nonumber\\
    &\;\;\;\;\;\;
        +\int_0^t\e^{-\int_\tau^ts_i(r)dr}h_{i+1}(\tau)d\tau,
        \;\; \forall t\in[0,t_f).
\end{flalign}

(i) For the second term of (\ref{eq:zi-equal}), note that
\begin{flalign}
    \frac{|w_i|}{s_i}
    \leq\frac{|w_i|}{c_i+\kappa_i|w_i|^2}
    =\frac{1}{\frac{c_i}{|w_i|}+\kappa_i|w_i|}
    \leq\frac{1}{2\sqrt{c_i\kappa_i}}.
\end{flalign}
Hence, if $\tilde\theta\in L_\infty[0,t_f)$, we obtain
\begin{flalign}
    &\int_0^t\e^{-\int_\tau^ts_i(r)dr}|w_i(\tau)^T\tilde\theta(\tau)| d\tau
        \nonumber\\
    &\;\;\;\;\;\;
        \leq\|\tilde\theta\|_\infty\int_0^t\e^{-\int_\tau^ts_i(r)dr}s_i(\tau)\frac{|w_i(\tau)|}{s_i(\tau)} d\tau
        \nonumber\\
    &\;\;\;\;\;\;
        \leq\frac{\|\tilde\theta\|_\infty}{2\sqrt{c_i\kappa_i}}\int_0^t\e^{-\int_\tau^ts_i(r)dr}s_i(\tau) d\tau
        \nonumber\\
    &\;\;\;\;\;\;
        =\frac{\|\tilde\theta\|_\infty}{2\sqrt{c_i\kappa_i}}\e^{-\int_0^ts_i(r)dr}\int_0^t\e^{\int_0^\tau s_i(r)dr}s_i(\tau)d\tau
        \nonumber\\
    &\;\;\;\;\;\;
        =\frac{\|\tilde\theta\|_\infty}{2\sqrt{c_i\kappa_i}}\e^{-\int_0^ts_i(r)dr}\int_0^t\e^{\int_0^\tau s_i(r)dr}d\bigg(\int_0^\tau s_i(r)dr\bigg)
        \nonumber\\
    &\;\;\;\;\;\;
        =\frac{\|\tilde\theta\|_\infty}{2\sqrt{c_i\kappa_i}}\bigg(1-\e^{-\int_0^ts_i(r)dr}\bigg)
        \nonumber\\
    &\;\;\;\;\;\;
        \leq\frac{\|\tilde\theta\|_\infty}{2\sqrt{c_i\kappa_i}}
        \label{eq:zi-equal-final-term-A}
\end{flalign}
Similarly, because
\begin{flalign}
    \frac{1}{s_i}\bigg|\frac{\partial\alpha_{i-1}}{\partial\hat\theta}\bigg|^2\leq\frac{1}{2\sqrt{c_ig_i}},
\end{flalign}
it follows that, if $\dot{\hat\theta}\in L_\infty[0,t_f)$,
\begin{flalign}\label{eq:zi-equal-final-term-B}
    \int_0^t\e^{-\int_\tau^ts_i(r)dr}\bigg|\frac{\partial\alpha_{i-1}}{\partial\hat\theta}\dot{\hat\theta}\bigg| d\tau
    &\leq\frac{\|\dot{\hat\theta}\|_\infty}{2\sqrt{c_ig_i}}\int_0^t\e^{-\int_\tau^ts_i(r)dr}s_i(\tau) d\tau
        \nonumber\\
    &\leq
    \frac{\|\dot{\hat\theta}\|_\infty}{2\sqrt{c_ig_i}}
\end{flalign}
Combining (\ref{eq:zi-equal}), (\ref{eq:zi-equal-final-term-A}) and (\ref{eq:zi-equal-final-term-B}), we have
\begin{flalign}
    |h_i(t)|
    &\leq |h_i(0)|\e^{-\int_0^ts_i(r)dr}
        +\frac{\|\tilde\theta\|_\infty}{2\sqrt{{\underline{c}}\,{\underline{\kappa}}}}
        +\frac{\|\dot{\hat\theta}\|_\infty}{2\sqrt{\underline{c}\,\underline{g}}}
        \nonumber\\
    &\;\;\;\;
        +\int_0^t\e^{-\int_\tau^t s_i(r)dr}|h_{i+1}(\tau)|d\tau
        \nonumber\\
    &\leq |h_i(0)|\e^{-\underline{c}t}
        +\frac{\|\tilde\theta\|_\infty}{2\sqrt{{\underline{c}}\,{\underline{\kappa}}}}
        +\frac{\|\dot{\hat\theta}\|_\infty}{2\sqrt{\underline{c}\,\underline{g}}}
        \nonumber\\
    &\;\;\;\;
        +\int_0^t\e^{-\underline{c}(t-\tau)}|h_{i+1}(\tau)|d\tau,
        \;\; i=1,\ldots,n.
        \label{eq:after-invar-const-fomla-comb-A}
\end{flalign}
Now we are ready to prove by induction that $h_i(t)$ satisfies
\begin{equation}\label{eq-L-infty-hi-abs}
    |h_i(t)|\leq\e^{-\underline{c}t}\sum_{p=i}^n\frac{t^{p-i}}{(p-i)!}|h_p(0)|+h_i^*,\;\;i=1\ldots,n
\end{equation}
where
\begin{equation}
    h_i^*=\frac{\underline{c}^{n-i+1}-1}{\underline{c}^{n-i}(\underline{c}-1)}
    \Bigg(\frac{\|\tilde\theta\|_\infty}{2\sqrt{\underline{c}\,\underline{\kappa}}}+\frac{\|\dot{\hat\theta}\|_\infty}{2\sqrt{\underline{c}\,\underline{g}}}\Bigg).
\end{equation}
Recalling $h_{n+1}=u-\bar{u}=0$, it follows from (\ref{eq:after-invar-const-fomla-comb-A}) that $h_n(t)$ satisfies (\ref{eq-L-infty-hi-abs}). For $k=1,\ldots,n-1$, assume that $h_{k+1}(t)$ satisfies (\ref{eq-L-infty-hi-abs}), i.e.,
\begin{equation}\label{eq-L-infty-hi-abs-induction-proof-k-plus}
    |h_{k+1}(t)|\leq\e^{-\underline{c}t}\sum_{p=k+1}^n\frac{t^{p-(k+1)}}{[p-(k+1)]!}|h_p(0)|+h_{k+1}^*.
\end{equation}
We next show that $h_k(t)$ also satisfies (\ref{eq-L-infty-hi-abs}). By combining (\ref{eq:after-invar-const-fomla-comb-A}) and (\ref{eq-L-infty-hi-abs-induction-proof-k-plus}), we have
\begin{flalign}
    |h_k(t)|
    &\leq |h_k(0)|\e^{-\underline{c}t}
        +\frac{\|\tilde\theta\|_\infty}{2\sqrt{{\underline{c}}\,{\underline{\kappa}}}}
        +\frac{\|\dot{\hat\theta}\|_\infty}{2\sqrt{\underline{c}\,\underline{g}}}
        \nonumber\\
    &\;\;\;\;
        +\e^{-\underline{c}t}\sum_{p=k+1}^n\frac{|h_p(0)|}{[p-(k+1)]!}\int_0^t\tau^{p-(k+1)}d\tau
        \nonumber\\
    &\;\;\;\;
        +h_{k+1}^*\int_0^t\e^{-\underline{c}(t-\tau)}d\tau
        \nonumber\\
    &\leq |h_k(0)|\e^{-\underline{c}t}
        + \e^{-\underline{c}t}\sum_{p=k+1}^n\frac{t^{p-k}}{(p-k)!}|h_p(0)|
        \nonumber\\
    &\;\;\;\;
        +\frac{\|\tilde\theta\|_\infty}{2\sqrt{{\underline{c}}\,{\underline{\kappa}}}}
        +\frac{\|\dot{\hat\theta}\|_\infty}{2\sqrt{\underline{c}\,\underline{g}}}
        +\frac{h_{k+1}^*}{\underline{c}}
        \nonumber\\
    &=\e^{-\underline{c}t}\sum_{p=k}^n\frac{t^{p-k}}{(p-k)!}|h_p(0)|+h_k^*.
\end{flalign}
Hence, $h_i(t)$ satisfies (\ref{eq-L-infty-hi-abs}) for all $i=1,\ldots,n$. This implies that $h=[h_1,\ldots,h_n]^\T$ belongs to $L_\infty[0,t_f)$. Since the change of coordinates (\ref{eq:coordinate-change-A})-(\ref{eq:coordinate-change-B}) is smooth in $x$ and $\hat\theta$, and is bounded with respect to $t$, it follows that $x\in L_\infty[0,t_f)$. For any $h_1(0)=y(0)-r(0)\geq0$, the choice of $c_1$, \ldots, $c_n$ in (\ref{eq:choice-of-ci}) ensures that $h_2(0)$, \ldots, $h_n(0)\geq0$. Then, from (\ref{eq-L-infty-hi-abs}), we obtain
\begin{flalign}
    h_1(t)\leq|h_1(t)|\leq\e^{-\underline{c}t}\sum_{p=1}^n\frac{t^{p-1}}{(p-1)!}h_p(0)+h_1^*
\end{flalign}
which proves the second inequality of (\ref{eq:zi-L_infty-bd}). Moreover, since $h_2(0)$, \ldots, $h_n(0)\geq0$ and $h\in L_\infty[0,t_f)$, it follows from (\ref{eq:zi-equal}), (\ref{eq:zi-equal-final-term-A}) and (\ref{eq:zi-equal-final-term-B}) that
\begin{flalign}
    h_i(t)
    &\geq
        -\frac{\|\tilde\theta\|_\infty}{2\sqrt{{\underline{c}}\,{\underline{\kappa}}}}
        -\frac{\|\dot{\hat\theta}\|_\infty}{2\sqrt{\underline{c}\,\underline{g}}}
        \nonumber\\
    &\;\;\;\;
        +\frac{1}{c_i}\inf_{t\in[0,t_f)}\min\{h_{i+1}(t),0\},
        \; i=1,\ldots,n.
\end{flalign}
Hence,
\begin{flalign}
    h_{n-1}(t)
    \geq-\frac{\underline{c}^{2}-1}{\underline{c}(\underline{c}-1)}\Bigg(\frac{\|\tilde\theta\|_\infty}{2\sqrt{\underline{c}\,\underline{\kappa}}}+\frac{\|\dot{\hat\theta}\|_\infty}{2\sqrt{\underline{c}\,\underline{g}}}\Bigg).
\end{flalign}
By induction, we establish the first inequality of (\ref{eq:zi-L_infty-bd}).

(ii)
With the Cauchy-Schwarz inequality and in view of (\ref{eq:si}), we obtain that, if $\dot{\hat\theta}\in L_2[0,t_f)$,
\begin{flalign}
    &\int_0^t\e^{\int_0^\tau s_i(r)dr}\bigg|\frac{\partial\alpha_{i-1}}{\partial\hat\theta}\bigg||\dot{\hat\theta}| d\tau
        \nonumber\\
    &\leq
    \sqrt{\int_0^t|\dot{\hat\theta}|^2d\tau}\sqrt{\int_0^t\e^{2\int_0^\tau s_i(r)dr}\Big|\frac{\partial\alpha_{i-1}}{\partial\hat\theta}\Big|^2d\tau}
        \nonumber\\
    &\leq\|\dot{\hat\theta}\|_2\sqrt{\frac{1}{g_i}\int_0^t\e^{2\int_0^\tau s_i(r)dr}s_i(\tau)d\tau}
        \nonumber\\
    &=\|\dot{\hat\theta}\|_2\sqrt{\frac{1}{2g_i}\Big(\e^{2\int_0^t s_i(r)dr}-1\Big)}
        \nonumber\\
    &\leq\frac{\|\dot{\hat\theta}\|_2}{\sqrt{2g_i}}\e^{\int_0^t s_i(r)dr}.
\end{flalign}
Hence,
\begin{flalign}\label{eq:zi-equal-final-term-dotTheta-L2}
    \int_0^t\e^{-\int_\tau^ts_i(r)dr}\bigg|\frac{\partial\alpha_{i-1}}{\partial\hat\theta}\dot{\hat\theta}\bigg| d\tau
    \leq
    \frac{\|\dot{\hat\theta}\|_2}{\sqrt{2g_i}}.
\end{flalign}
With (\ref{eq:zi-equal}), (\ref{eq:zi-equal-final-term-A}) and (\ref{eq:zi-equal-final-term-dotTheta-L2}), we have
\begin{flalign}
    |h_i(t)|
    &\leq |h_i(0)|\e^{-\int_0^ts_i(r)dr}
        +\frac{\|\tilde\theta\|_\infty}{2\sqrt{{\underline{c}}\,{\underline{\kappa}}}}
        +\frac{\|\dot{\hat\theta}\|_2}{\sqrt{2\underline{g}}}
        \nonumber\\
    &\;\;\;\;
        +\int_0^t\e^{-\int_\tau^t s_i(r)dr}|h_{i+1}(\tau)|d\tau
        \nonumber\\
    &\leq |h_i(0)|\e^{-\underline{c}t}
        +\frac{\|\tilde\theta\|_\infty}{2\sqrt{{\underline{c}}\,{\underline{\kappa}}}}
        +\frac{\|\dot{\hat\theta}\|_2}{\sqrt{2\underline{g}}}
        \nonumber\\
    &\;\;\;\;
        +\int_0^t\e^{-\underline{c}(t-\tau)}|h_{i+1}(\tau)|d\tau,
        \;\; i=1,\ldots,n
\end{flalign}
which is identical to (\ref{eq:after-invar-const-fomla-comb-A}), except that $\|\dot{\hat\theta}\|_\infty$ is replaced by $\|\dot{\hat\theta}\|_2$. Hence, by induction, one can show that $h_i(t)$ satisfies (\ref{eq-L-infty-hi-abs}) with
\begin{flalign}
    h_i^*=\frac{\underline{c}^{n-i+1}-1}{\underline{c}^{n-i}(\underline{c}-1)}
    \scalebox{1.1}{\Bigg(}\frac{\|\tilde\theta\|_\infty}{2\sqrt{\underline{c}\,\underline{\kappa}}}+\frac{\|\dot{\hat\theta}\|_2}{\sqrt{2\underline{g}}}\scalebox{1.1}{\Bigg)}.
\end{flalign}
Repeating the same argument used in the proof of property (i), we establish that $h,x \in L_\infty[0,t_f)$, and $h_1(t)$ satisfies (\ref{eq:zi-L_infty-bd}) with $h_1^*$ given in (\ref{eq:h-star-bd-L2}).
\hfill $\Box$
\vskip5pt

\begin{remark}\em
Although adaptive safety-critical control has been studied in \cite{taylor2020adaptive, lopez2020robust, isaly2021adaptive, wang2024adaptive}, these results are restricted by the relative-degree-one safety constraints. For high-relative-degree constraints, a recent attempt is made in \cite{cohen2023modular} for a nonlinear affine control system of the form $\dot{x}=f(x)+F(x)\theta+g(x)u$, under the assumption that the derivative of a CBF candidate $h_1(x)$ satisfies
\begin{flalign}
    &\dot{h}_i=-c_ih_i+h_{i+1},\;\;i=1,\ldots,n-1
        \nonumber\\
    &\dot{h}_n=L_fh_n+L_gh_nu+L_Fh_n\theta,
        \label{eq:cohen-acbf-cond}
\end{flalign}
which implies that unknown parameter $\theta$ and the control $u$ are required to be matched, i.e., $\theta$ must lie in the span of $u$. In fact, by taking a new CBF candidate $\hbar(x)={h}_n(x)$, this problem reduces to the relative-degree-one problems studied in \cite{taylor2020adaptive, lopez2020robust, isaly2021adaptive, wang2024adaptive}. In contrast, our problem is not restricted by a matching condition. Moreover, as shown (\ref{eq:dot-zi-final-state-fb}), when the uncertainty is not matched and the CBF candidate $h_1(x)$ is differentiated twice or more, the higher derivatives of $h_1(x)$ is influenced by not only the parameter estimation error $\tilde{\theta}$ but also its derivative $\dot{\tilde\theta}=-\dot{\hat\theta}$. This is different from all the relative-degree-one results \cite{taylor2020adaptive, lopez2020robust, isaly2021adaptive, wang2024adaptive}. Consequently, our problem cannot be addressed by a relative-degree-one ACBF approach.
\end{remark}

\section{Error-observer Schemes for Parameter Identification}\label{eq:z-passive-scheme}

In this section, we present two parameter identification schemes based on the information of the error system model (\ref{eq:dot-z-A}) to ensure that the pair $(\tilde{\theta},\dot{\hat\theta})$ satisfies the requirement in Lemma \ref{lem:ISSf-state-feedback}.

\subsection{$h$-Passive Identifier}

Introduce the observer
\begin{flalign}\label{eq:dot-hat-z-state-fb}
    &\dot{\hat{h}}=A(h,\hat\theta,t)\hat{h}+\sigma W(h,\hat\theta,t)^\T
        \nonumber\\
    &\;\;\;\;\;\;\;\;\times W(h,\hat\theta,t)P(h-\hat{h})+Q(h,\hat\theta,t)^\T\dot{\hat\theta}
\end{flalign}
where $\sigma>0$ is a design parameter, $A$, $W$ and $Q$ are matrix-valued functions defined in (\ref{eq:def-A})-(\ref{eq:def-Q}), respectively, and $P=P^T>0$ is the solution to the Lyapunov equation
\begin{flalign}\label{eq:Ac-lya-eq}
    A_0^\T P+PA_0=-I
\end{flalign}
with
\begin{flalign}\label{eq:def-matrix-AC}
    A_0
    =
    \left[
      \begin{array}{cccc}
        -c_1 & 1 & \cdots & 0 \\
        0   & -c_2 & \ddots & \vdots \\
        \vdots & \ddots & \ddots & 1 \\
        0   & \cdots & 0 & -c_n \\
      \end{array}
    \right].
\end{flalign}
By taking $u=\bar{u}$, it follows from (\ref{eq:dot-z-A}) and (\ref{eq:dot-hat-z-state-fb}) that the observer error
\begin{flalign}\label{eq:z-passive-epsilon}
    \epsilon=h-\hat{h}
\end{flalign}
is governed by
\begin{flalign}\label{eq:dot-tilde-z}
    \dot{\epsilon}
    &=A(h,\hat\theta,t)\epsilon
        -\sigma W(h,\hat\theta,t)^\T
        \nonumber\\
    &\;\;\;\;\;\;\;\;
        \times W(h,\hat\theta,t)P\epsilon
        +W(h,\hat\theta,t)^\T\tilde\theta.
\end{flalign}
Consider the Lyapunov function candidate
\begin{flalign}\label{eq:Lya-identifier}
    V=\epsilon^\T P\epsilon
        +\tilde{\theta}^\T\Gamma^{-1}\tilde\theta,\;\;\Gamma=\Gamma^\T>0.
\end{flalign}
With (\ref{eq:Ac-lya-eq}) and (\ref{eq:dot-tilde-z}), we get
\begin{flalign}
    \dot{V}
    &=\epsilon^\T\big(A^\T P+PA\big)\epsilon
        -2\sigma\epsilon^\T PW^\T W P\epsilon
        \nonumber\\
    &\;\;\;\;\;\;\;\;\;\;\;\;
        +2\epsilon^\T PW^\T\tilde\theta
        -2\tilde{\theta}^\T\Gamma^{-1}\dot{\hat{\theta}}
        \nonumber\\
    &=\epsilon^\T\big(A_0^\T P+PA_0\big)\epsilon
        +2\epsilon^\T\big(A-A_0\big)P\epsilon
        \nonumber\\
    &\;\;\;\;\;\;\;\;\;\;\;\;
        -2\sigma|WP\epsilon|^2
        +2\epsilon^\T PW^\T\tilde\theta
        -2\tilde{\theta}^\T\Gamma^{-1}\dot{\hat{\theta}}
        \nonumber\\
    &\leq-|\epsilon|^2
        -2\sigma|WP\epsilon|^2
        +2\tilde\theta^\T WP\epsilon
        -2\tilde{\theta}^\T\Gamma^{-1}\dot{\hat{\theta}}
        \label{eq:dot-V-state-fb}
\end{flalign}
Design the update law for $\hat\theta$ as
\begin{flalign}
    \dot{\hat\theta}
    =\Gamma WP\epsilon,\;\;\;\;\Gamma=\Gamma^\T>0.
    \label{eq:dot-tilde-theta}
\end{flalign}
Then we have
\begin{flalign}\label{eq:dot-V-bd-ci-si}
    \dot{V}
    \leq-|\epsilon|^2
        -2\sigma|WP\epsilon|^2.
\end{flalign}

\begin{lemma}\label{lem:adaptive-bd-z-passive}
Let $V$ be a continuously differentiable Lyapunov function as defined in (\ref{eq:Lya-identifier}). Suppose that the solutions of (\ref{eq:dot-z-A}), (\ref{eq:dot-tilde-z}), and (\ref{eq:dot-tilde-theta}) are defined on $[0,t_f)$. Then the following properties hold:
\begin{itemize}
  \item[(i)] $\tilde{\theta}\in L_\infty[0,t_f)$.
  \item[(ii)] $\epsilon\in L_2[0,t_f)\bigcap L_\infty[0,t_f)$.
  \item[(iii)] $\dot{\hat{\theta}}\in L_2[0,t_f)$.
  \item[(iv)] $\|\tilde\theta\|_\infty\leq\sqrt{\bar\lambda(\Gamma)V(0)}$ and $\|\dot{\hat\theta}\|_2\leq\frac{\bar\lambda(\Gamma)}{\sqrt{2\sigma}}\sqrt{V(0)}$. Moreover, for the case of $\hat h(0)=h(0)$ and $\Gamma=\gamma I$, $\|\tilde\theta\|_\infty\leq|\tilde\theta(0)|$
      and
      $\|\dot{\hat\theta}\|_2\leq\sqrt{\frac{\gamma}{2\sigma}}|\tilde\theta(0)|$.
\end{itemize}
\end{lemma}

\noindent
\textbf{Proof.}
The nonpositivity of $\dot{V}$ implies $\epsilon, \tilde\theta\in L_\infty[0,t_f)$. From (\ref{eq:dot-V-bd-ci-si}),
\begin{equation}
    \int_0^t|\epsilon(\tau)|^2d\tau
    \leq V(0)-V(t)\leq V(0),
\end{equation}
which implies $\epsilon\in L_2[0,t_f)$. From (\ref{eq:dot-tilde-theta}),
\begin{equation}\label{eq:dot-hatTheta-sq-bd}
    |\dot{\hat\theta}|^2
        \leq\bar{\lambda}(\Gamma)^2|WP\epsilon|^2.
\end{equation}
Combining this with (\ref{eq:dot-V-bd-ci-si}) yields
\begin{equation}\label{eq:L2-bd-dot-theta}
    \|\dot{\hat\theta}\|_2^2
    \leq \frac{\bar{\lambda}(\Gamma)^2}{2\sigma}(V(0)-V(t))
    \leq \frac{\bar{\lambda}(\Gamma)^2}{2\sigma}V(0),
\end{equation}
which concludes property (iii). Moreover, from
$\tilde\theta(0)^\T\Gamma\tilde\theta(0)\leq V(0)$
and (\ref{eq:L2-bd-dot-theta}), property (iv) holds.
\hfill $\Box$
\vskip5pt

Combining Lemma \ref{lem:ISSf-state-feedback} (ii) and Lemma \ref{lem:adaptive-bd-z-passive}, we have the following result.

\begin{theorem}[$h$-Passive]\label{thm:adaptive-state-fb}
For the closed-loop adaptive system consisting of the plant (\ref{eq:str-fb-sys}), controller (\ref{eq:state-fb-controller}), and parameter identifier (\ref{eq:dot-hat-z-state-fb}), (\ref{eq:dot-tilde-theta}), the following properties hold:
\begin{itemize}
\item[(i)] All signals in the closed-loop system are globally uniformly bounded, and globally asymptotic tracking is achieved.
\item[(ii)] For choice of $c_1$, \ldots, $c_n$ satisfying (\ref{eq:choice-of-ci}), if $h_1(0)\geq0$, $\hat h(0)=h(0)$, and $\Gamma=\gamma I$, then $h_1(t)$ satisfies (\ref{eq:zi-L_infty-bd}) with
\begin{flalign}\label{eq:passive-cls-loop-viol-bd}
    h_1^*
    =\frac{\underline{c}^{n}-1}{2\underline{c}^{n-1}(\underline{c}-1)}
    \Bigg(\frac{1}{\sqrt{\underline{c}\,\underline{\kappa}}}+\sqrt{\frac{\gamma}{\sigma\underline{g}}}\Bigg)|\tilde\theta(0)|.
\end{flalign}
\end{itemize}
\end{theorem}

\noindent
\textbf{Proof.}
Property (ii) can be established by combining Lemma \ref{lem:ISSf-state-feedback} (ii) and Lemma \ref{lem:adaptive-bd-z-passive} (iv). In the following, we prove property (i).

We first prove the boundedness of closed-loop signals. By combining Lemma \ref{lem:ISSf-state-feedback} (i) and Lemma \ref{lem:adaptive-bd-z-passive}, we get that $\tilde\theta$, $h$, $x$ and $\epsilon$ belong to $L_\infty[0,t_f)$. Recalling (\ref{eq:z-passive-epsilon}), it follows that $\hat{h}\in L_\infty[0,t_f)$. Hence, all closed-loop signals are bounded on $[0,t_f)$ by constants that are independent of $t_f$. By contradiction, we conclude that all closed-loop signals are bounded for all $t\geq0$.

Next, we show the convergence of $h_1(t)$. For convenience, let $\hat{h}_{n+1}=0$. From (\ref{eq:dot-hat-z-state-fb}),
\begin{flalign}\label{eq:z-passive-dot-hat-zi-sq}
    \frac{1}{2}\frac{d}{dt}\hat{h}_i^2
    &=-s_i\hat{h}_i^2
        +\hat{h}_{i}\hat{h}_{i+1}
        \nonumber\\
    &\;\;\;\;
        +\sigma\hat{h}_i[W^\T WP]_{(i)}\epsilon
        -\hat{h}_i\frac{\partial\alpha_{i-1}}{\partial\hat\theta}\dot{\hat\theta}
        \nonumber\\
    &\leq-\frac{c_i}{4}\hat{h}_i^2
            -\bigg(\frac{c_i}{4}\hat{h}_i^2
            -\hat{h}_i\hat{h}_{i+1}\bigg)
            \nonumber\\
    &\;\;\;\;
            -\Bigg(g_i\bigg|\frac{\partial\alpha_{i-1}}{\partial\hat\theta}\hat{h}_i\bigg|^2
            -\bigg|\frac{\partial\alpha_{i-1}}{\partial\hat\theta}\hat{h}_i\bigg||\dot{\hat\theta}|\Bigg)
        \nonumber\\
    &\;\;\;\;
            -\bigg(\frac{c_i}{2}\hat{h}_i^2
            -\sigma|\hat{h}_i||W^\T WP|_{2}|\epsilon|\bigg)
        \nonumber\\
    &\leq-\frac{c_i}{4}\hat{h}_i^2
            +\frac{1}{c_i}\hat{h}_{i+1}^2
            +\frac{1}{4g_i}|\dot{\hat\theta}|^2
            +\frac{\sigma^2}{2c_i}|W^\T WP|_2^2|\epsilon|^2.
\end{flalign}
Because $x$ is bounded and $\alpha_i$ is smooth, it follows from (\ref{eq:wi}) and (\ref{eq:def-W}) that $W$ is bounded. From Lemma \ref{lem:adaptive-bd-z-passive}, we have $\dot{\hat\theta},\epsilon\in L_2$. Therefore, by applying \cite[Lemma B.5]{krstic1995nonlinear} to (\ref{eq:z-passive-dot-hat-zi-sq}), we obtain $\hat{h}_n\in L_2$. Repeating this argument iteratively yields $\hat{h}_1,\ldots,\hat{h}_{n-1}\in L_2$, and thus, $\hat{h}\in L_2$. It then follows that $h=\epsilon+\hat{h}\in L_2$. By Barbalat's lemma (see, e.g., \cite[Corollary A.7]{krstic1995nonlinear}), $h(t)\rightarrow0$ as $t\rightarrow+\infty$, which establishes the convergence of $h_1(t)$.
\hfill $\Box$
\vskip5pt

\begin{remark}\em
As in Lemma \ref{lem:ISSf-state-feedback} (ii) and Lemma \ref{lem:adaptive-bd-z-passive} (iv), choosing initialization $\hat{h}(0)=h(0)$ for the observer (\ref{eq:dot-hat-z-state-fb}) can eliminate the influence of the initial observer error $\epsilon(0)$ on the safety violation bound $h_1^*$. Under this observer initialization, together with the nonovershooting backsteping control (\ref{eq:alpha-0})-(\ref{eq:state-fb-controller}) and the $h$-passive parameter identifier (\ref{eq:dot-hat-z-state-fb}), the safety violation bound $h_1^*$ depends on the design parameters $\underline{c}$, $\underline{\kappa}$, $\underline{g}$, $\sigma$, and $\gamma$, as well as the initial parameter estimation error $\tilde{\theta}(0)$. As shown in (\ref{eq:passive-cls-loop-viol-bd}), the maximal overshoot $h_1^*$ can be reduced by increasing $\underline{c}$, $\underline{\kappa}$ and $\underline{g}$. Furthermore, by increasing $\underline{c}$ appropriately, we can make this bound arbitrarily small.
\end{remark}

\begin{remark}\em
The safety violation bound $h_1^*$ estimated in (\ref{eq:passive-cls-loop-viol-bd}) may be conservative for some systems.
\begin{itemize}
  \item First, increasing the adaptation gain $\gamma$ can accelerate the adaptation. In most cases, the relationship between $h_1^*$ and $\gamma$ does not match (\ref{eq:passive-cls-loop-viol-bd}), especially when $\hat{\theta}(t)$ does not cross the the true value $\hat\theta(t)=\theta$ and oscillate around it; for instance, see Example \ref{example:z-passive}.
  \item Second, (\ref{eq:passive-cls-loop-viol-bd}) estimates $h_1^*$ using $V(0)$, the value of $V(t)$ at system initialization. However, since $V(t)$ is non-increasing and the state $x(t)$ must cross the boundary of the safe set $S = \{x : h_1(x, t) \geq 0\}$ before any safety violation occurs, a less conservative estimate of $h_1^*$ may be obtained by using $V(t_1)$ [even though $V(t_1)$ is not available a priori], where $t_1=\inf\{t\geq0:h_1(x(t,x_0),t)=0\}$ is the first time the state trajectory $x(t,x_0)$ reaches the boundary of $S$.
\end{itemize}
\end{remark}

\subsection{$h$-Swapping Identifier}

Since the $h$-passive parameter identifier cannot guarantee the boundedness of $\dot{\hat\theta}$. This section develops a swapping scheme, modified from \cite{krstic1995nonlinear}, to ensure that $\dot{\hat\theta}$ is both bounded and square-integrable.

Introduce the filters
\begin{flalign}
    \dot{\Omega}^\T
    &=A(h,\hat\theta,t)\Omega^\T+W(h,\hat\theta,t)^\T,\;\;\;\;\Omega\in\R^{p\times n}
        \label{eq:big-omega}\\
    \dot{\Omega}_0
    &=A(h,\hat\theta,t)\Omega_0+W(h,\hat\theta,t)^\T\hat\theta-Q(z,\hat\theta,t)^\T\dot{\hat\theta},\;\;\Omega_0\in\R^n
        \label{eq:big-omega0}
\end{flalign}
where $A$, $W$ and $Q$ are matrix-valued functions defined in (\ref{eq:def-A})-(\ref{eq:def-Q}), respectively. Let
\begin{flalign}
    \epsilon=h+\Omega_0-\Omega^\T\hat\theta.
    \label{eq:swapping-epsilon}
\end{flalign}
With \cite[Lemma F.1]{krstic1995nonlinear}, we establish
\begin{flalign}\label{eq:z-swap-estim-error}
    \epsilon=\Omega^\T\tilde\theta+\tilde\epsilon
\end{flalign}
where $\tilde\epsilon$ is governed by
\begin{flalign}\label{eq:swap-state-fbdot-tilde-epsi}
    \dot{\tilde\epsilon}=A(h,\hat\theta,t)\tilde\epsilon.
\end{flalign}
Design the update law for $\hat\theta$ as
\begin{flalign}
    \dot{\hat\theta}=\Gamma\frac{\Omega\epsilon}{1+\nu|\Omega|_F^2},\;\;\;\;\Gamma=\Gamma^\T>0,\;\;\nu\geq0
    \label{eq:hat-theta}
\end{flalign}

\begin{lemma}\label{lem:bd-adaptive-signal}
Let $V=\tilde{\epsilon}^TP\tilde\epsilon+\tilde\theta^T\Gamma^{-1}\tilde\theta$. Suppose that the solutions of (\ref{eq:big-omega}), (\ref{eq:big-omega0}) and (\ref{eq:hat-theta}) are defined on $[0,t_f)$. Then the following properties hold:
\begin{itemize}
\item[(i)] $\tilde{\theta}\in L_\infty[0,t_f)$.
\item[(ii)] $\epsilon\in L_2[0,t_f)\bigcap L_\infty[0,t_f)$.
\item[(iii)] $\dot{\hat\theta}\in L_2[0,t_f)\bigcap L_\infty[0,t_f)$.
\item[(iv)] $\|\tilde\theta\|_\infty\leq\sqrt{\bar\lambda(\Gamma)V(0)}$ and $\|\dot{\hat\theta}\|_\infty\leq\frac{\bar\lambda(\Gamma)}{\nu}|\tilde\theta(0)|$. Moreover, for the case of $\Omega(0)=0$, $\Omega_0(0)=-h(0)$ and $\Gamma = \gamma I$, $\|\tilde\theta\|_\infty\leq|\tilde\theta(0)|$.
\end{itemize}
\end{lemma}

\noindent
\textbf{Proof.} Compared to the proof of \cite[Lemma 6.4]{krstic1995nonlinear}, the only difference lies in proving the boundedness of $\Omega$. For $j=1,\ldots,p$ and $i=1,\ldots,n$, it follows from (\ref{eq:big-omega}) that the $(j,i)$ element of $\Omega$ satisfeis
\begin{flalign}
    \dot{\Omega}_{j,i}=-s_i\Omega_{j,i}+\Omega_{j,i+1}+W_{j,i}.
\end{flalign}
Hence,
\begin{flalign}
    \frac{1}{2}\frac{d}{dt}\Omega_{j,i}^2
    &=-s_i\Omega_{j,i}^2+\Omega_{j,i}\Omega_{j,i+1}+\Omega_{j,i}W_{j,i}
        \nonumber\\
    &\leq-s_i\Omega_{j,i}^2+\Omega_{j,i}\Omega_{j,i+1}+|\Omega_{j,i}||w_i|
        \nonumber\\
    &\leq-c_i\Omega_{j,i}^2+\Omega_{j,i}\Omega_{j,i+1}+\frac{1}{4\kappa_i}.
\end{flalign}
Since $\Omega_{j,n+1}=0$, $\Omega_{j,n}$ is bounded, and thus,
\begin{flalign}
    \frac{1}{2}\frac{d}{dt}\Omega_{j,n-1}^2
    \leq-c_{n-1}\Omega_{j,n-1}^2+\|\Omega_{j,n}\|_\infty|\Omega_{j,n-1}|+\frac{1}{4\kappa_{n-1}},
\end{flalign}
which implies
\begin{flalign}
    &|\Omega_{j,n-1}|\geq\frac{\|\Omega_{j,n}\|_\infty}{2c_{n-1}}+\sqrt{\frac{\|\Omega_{j,n}\|_\infty^2}{4c_{n-1}^2}+\frac{1}{4c_{n-1}\kappa_{n-1}}}
        \nonumber\\
    &
    \Rightarrow\frac{1}{2}\frac{d}{dt}\Omega_{j,i}^2\leq0.
\end{flalign}
Hence,
\begin{flalign}
    |\Omega_{j,n-1}|
    &\leq|\Omega_{j,n-1}(0)|+\frac{\|\Omega_{j,n}\|_\infty}{2c_{n-1}}
        \nonumber\\
    &\;\;\;\;
    +\sqrt{\frac{\|\Omega_{j,n}\|_\infty^2}{4c_{n-1}^2}+\frac{1}{4c_{n-1}\kappa_{n-1}}}.
\end{flalign}
Therefore, $\Omega_{j,n-1}$ is bounded. Similarly, we obtain that $\Omega_{j,1}$, \ldots, $\Omega_{j,n-2}$ are bounded. Consequently, $\Omega$ is bounded.
\hfill $\Box$
\vskip5pt

Similar to Theorem \ref{thm:adaptive-state-fb}, the observer (\ref{eq:big-omega}), (\ref{eq:big-omega0}) can be initialized with $\Omega(0)=0$ and $\Omega_0(0)=-h(0)$ to eliminate the effect of the initial observer error and thereby reduce the safety violation bound. Combining Lemma \ref{lem:ISSf-state-feedback} (i) and Lemma \ref{lem:bd-adaptive-signal}, we have the following result.

\begin{theorem}[$h$-Swapping]
For the closed-loop adaptive system consisting of the plant (\ref{eq:str-fb-sys}), controller (\ref{eq:state-fb-controller}), filters (\ref{eq:big-omega}), (\ref{eq:big-omega0}), and update law (\ref{eq:hat-theta}), the following properties hold:
\begin{itemize}
\item[(i)] All signals in the closed-loop system are globally uniformly bounded, and globally asymptotic tracking is achieved.
\item[(ii)] For choice of $c_1$, \ldots, $c_n$ satisfying (\ref{eq:choice-of-ci}), if $h_1(0)\geq0$, $\Omega(0)=0$, $\Omega_0(0)=-h(0)$, and $\Gamma=\gamma I$, then $h_1(t)$ satisfies (\ref{eq:zi-L_infty-bd}) with
\begin{flalign}\label{eq:h-swap-h-star}
    h_1^*=
    \frac{\underline{c}^{n}-1}{2\underline{c}^{n-1}(\underline{c}-1)}
    \bigg(\frac{1}{\sqrt{\underline{c}\,\underline{\kappa}}}+\frac{\gamma}{\nu\sqrt{\underline{c}\,\underline{g}}}\bigg)|\tilde\theta(0)|.
\end{flalign}
\end{itemize}
\end{theorem}

\noindent
\textbf{Proof.}
Property (i) can be established with the proof of \cite[Theorem 6.3]{krstic1995nonlinear}, while property (ii) follows from the combination of Lemma \ref{lem:ISSf-state-feedback} (i) and Lemma \ref{lem:bd-adaptive-signal}.
\hfill $\Box$

\subsection{Application to Safety-Critical Tracking under Nonovershooting Constraints}

Now we are ready to show how the results of this section can be applied to minimally override the unsafe actions of the nominal controller under the high-relative-degree nonovershooting constraint. Due to space limitation, we only verify the effectiveness of Theorem \ref{thm:adaptive-state-fb}. From (\ref{eq:dot-z-A}), we see that any control satisfying $u\geq\bar{u}$ is safer than the nonovershooting override controller $\bar{u}$. Hence, for any nominal control $u_0$, the safety filter (\ref{general-safety-filter}) is rewritten as
\begin{equation}\label{eq:adaptive-QP}
    u=
    \mathop{\arg\min}_{u\in\R}|u-u_0|^2\;\;\;\;
    \text{s.t.}\;\; u\geq\bar{u}.
\end{equation}
As indicated in \cite{krstic2023inverse}, the solution to (\ref{eq:adaptive-QP}) is
\begin{flalign}
    u=\max\{\bar{u},u_0\},
\end{flalign}
which implies that: i) the override controller $\bar{u}$ is ``switched on'' to replace the nominal controller $u_0$ whenever $u_0$ is at the risk of safety violation in the sense of $u_0\leq\bar{u}$; and ii) the override controller is ``switched off'' when $u_0$ is sufficiently safe in the sense of $u_0>\bar{u}$. Since the parameter update law (\ref{eq:dot-tilde-theta}) is dependent on the error system (\ref{eq:dot-z-A}) driven by the override controller $\bar{u}$, the parameter estimation update should be switched off accordingly whenever the nominal controller is sufficiently safe. Hence, analogous to \cite{taylor2020adaptive}, the $h$-passive parameter update law (\ref{eq:dot-tilde-theta}) for $\hat{\theta}$ is modified as
\begin{flalign}\label{eq:example-z-passive-update-law}
    \dot{\hat\theta}=
    \left\{
      \begin{array}{ll}
        \gamma W(h,\hat\theta,t)P\epsilon, & \bar{u}\geq u_0 \\
        0, & \bar{u}\leq u_0
      \end{array}
    \right.
\end{flalign}

\begin{example}\label{example:z-passive}\em
Consider system
\begin{flalign}\label{eq:example-system}
    \dot{x}_1=x_2+\varphi_1(x_1)^\T\theta,\;\;\;\;
    \dot{x}_2=u+\varphi_2(x_1,x_2)^\T\theta
\end{flalign}
with
\begin{flalign}
    \varphi_1(x_1)=-8, \;\;\;\;\varphi_2(x_1,x_2)=-3, \;\;\;\;\theta=10.
\end{flalign}
The control objective is to force the output $y=x_1$ to asymptotically track the nominal reference trajectory $y_r(t)$, while ensuring that the safety constraint $y(t)\geq r(t)$ is satisfied for all $t\geq0$. The nominal controller $u_0$ is designed using the Lyapunov backstepping \cite{krstic1995nonlinear}, while the override controller $\bar{u}$ is constructed with the $h$-passive scheme in Theorem \ref{thm:adaptive-state-fb}. To avoid unnecessary complexity, we assume that $u_0$ knows the true value of $\theta$. Unless stated otherwise, all simulations are carried out with initial state $x(0)=(1.6, 84.5)$ and nominal values $\underline{c}=2.5$, $\underline{\kappa}=0.05$, $\underline{g}=0.3$, $\sigma=0.05$, and $\gamma=2$. In all simulations, we set $\hat{h}(0)=h(0)$, $c_1=c_2=\underline{c}$, $\kappa_1=\kappa_2=\underline{\kappa}$, and $g_2=\underline{g}$.

Figure \ref{Fig:z-passive-time-varying-boundary} illustrates the effectiveness of the $h$-passive scheme in Theorem \ref{thm:adaptive-state-fb}. It can be seen that the adaptation is able to reduce the parametric uncertainty. As time increases, the parameter estimate converges to the actual value, and the amount of safety violation decreases. However, we also see that the parameter estimate is not continuously differentiable. This is caused by the discontinuity in the adaptation law (\ref{eq:example-z-passive-update-law}).

Note that the adaptation law (\ref{eq:example-z-passive-update-law}) implies that adaptation is frozen whenever the nominal controller $u_0$ is safer than the override controller $\bar{u}$. This behavior can be seen in Figure \ref{Fig:z-passive-time-varying-boundary}, where the parameter estimate $\hat\theta(t)$ remains fixed at its initial value during the time interval $t \in [0, 0.8]$. In some cases, this phenomenon is detrimental. For instance, if the parameter estimate $\hat\theta(t)$ is poorly initialized, the override controller $\bar{u}$ may initially be unsafe. When the nominal controller $u_0$, although also unsafe, is safer than the override controller $\bar{u}$, the adaptation may be always turned off. As a result, the system may consistently execute the unsafe nominal command. To further illustrate this issue, we carried out another simulation in Figure \ref{Fig:z-passive-diff-hatTheta-init}. From the blue dash line, we see that when $\hat{\theta}(0)$ is selected improperly, the adaptation fails to activate to reduce the uncertainty, and the closed-loop system always executes the unsafe nominal control task. One possible solution to this issue is to carefully select the design parameters or the initial parameter estimate such that $\bar{u}$ is safer than $u_0$ at the system initialization. However, this approach comes at the cost of restricting the design flexibility in choosing the design parameters and initialization values.

Figure \ref{Fig:ZPassive-safety-violation-controller-para} and Figure \ref{Fig:zPassive-safety-violation-identifier-para} demonstrate the relationship between the safety violation and the design parameters. As illustrated in Figure \ref{Fig:ZPassive-safety-violation-controller-para}, increasing the controller parameters $\underline{c}$, $\underline{\kappa}$ and $\underline{g}$ in (\ref{eq:underline-c-kappa-g}) can reduce the safety violation bound. However, from Figure \ref{Fig:zPassive-safety-violation-identifier-para}, we see that the dependence of safety violation bound on the identifier parameters $\sigma$ in (\ref{eq:dot-hat-z-state-fb}) and $\gamma$ in (\ref{eq:example-z-passive-update-law}) does not match (\ref{eq:h-swap-h-star}). This implies that the bound estimated in (\ref{eq:h-swap-h-star}) may be conservative in certain cases.
\begin{figure}[!htb]
 \centering
 \includegraphics[width=6cm]{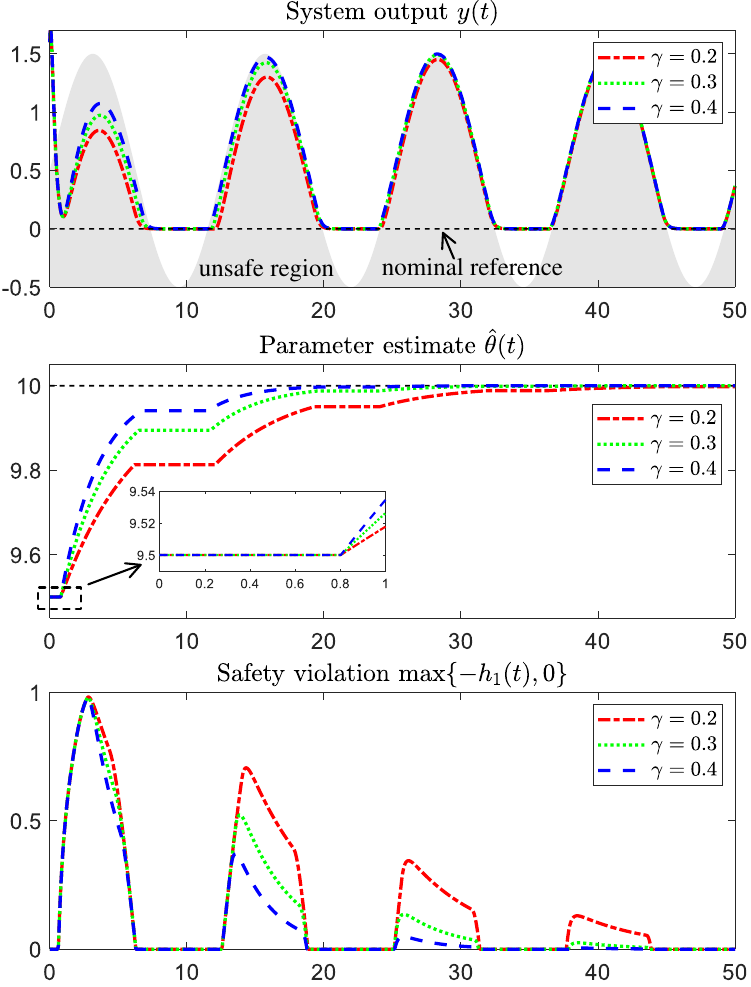}
 \caption{Responses of system (\ref{eq:example-system}) under the $h$-passive scheme with $\sigma=1$, $\hat\theta(0)=9.5$, $y_r(t)\equiv0$, and $r(t)=\sin(t/2) + 0.5$.}
\label{Fig:z-passive-time-varying-boundary}
\end{figure}
\begin{figure}[!htb]
 \centering
 \includegraphics[width=6cm]{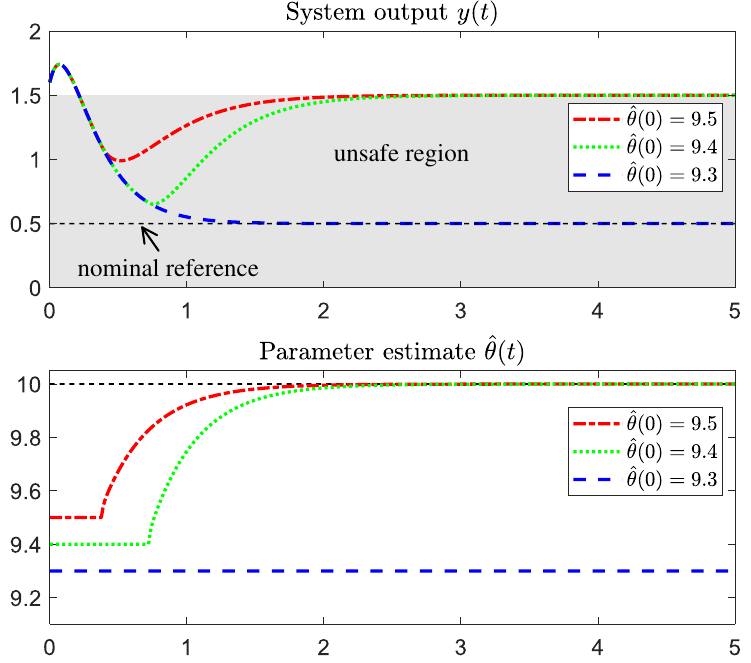}
 \caption{Effect of initial parameter estimate $\hat\theta(0)$ under the $h$-passive scheme with $y_r(t)\equiv0.5$ and $r(t)\equiv1.5$.}
\label{Fig:z-passive-diff-hatTheta-init}
\end{figure}
\begin{figure}[!htb]
 \centering
 \includegraphics[width=9cm]{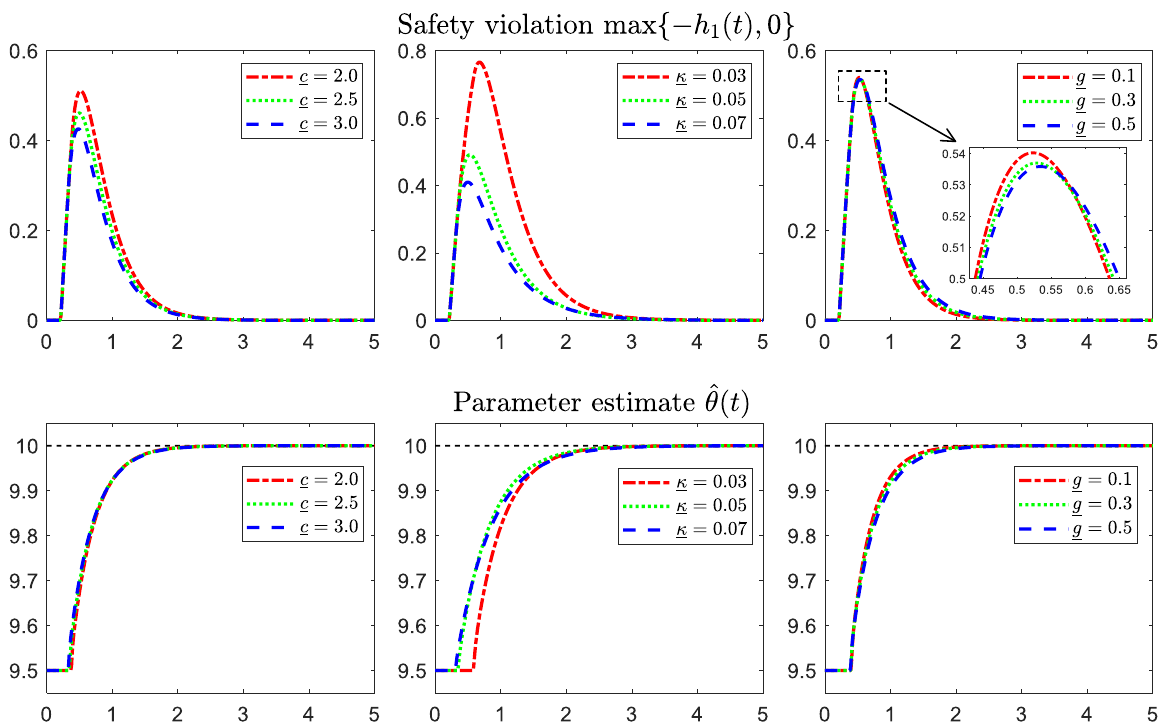}
 \caption{Dependence of safety violation on the controller parameters $\underline{c}$, $\underline{\kappa}$ and $\underline{g}$ in (\ref{eq:underline-c-kappa-g}) under the $h$-passive scheme with $\hat\theta(0)=9.5$, $y_r(t)\equiv0.5$, and $r(t)\equiv1.5$.}
\label{Fig:ZPassive-safety-violation-controller-para}
\end{figure}
\begin{figure}[!htb]
 \centering
 \includegraphics[width=6cm]{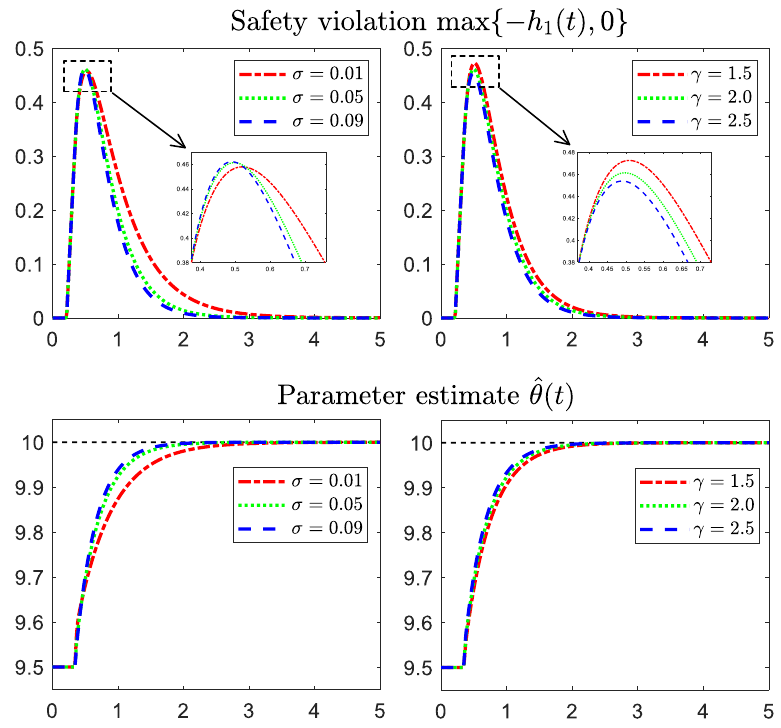}
 \caption{Dependence of safety violation on the identifier parameters $\sigma$ in (\ref{eq:dot-hat-z-state-fb}) and $\gamma$ in (\ref{eq:example-z-passive-update-law}) under the $h$-passive scheme with $\hat\theta(0)=9.5$, $y_r(t)\equiv0.5$, and $r(t)\equiv1.5$.}
\label{Fig:zPassive-safety-violation-identifier-para}
\end{figure}
\end{example}

\section{Plant-observer Schemes for Parameter Identification}\label{sec:x-passive}

As can be seen in the error-observer schemes above, the update laws depend on the backstepping coordinate change in (\ref{eq:coordinate-change-A})-(\ref{eq:coordinate-change-B}). However, different from the standard backstepping control\cite{krstic1995nonlinear}, the safety-critical control often involves the switching among multiple coordinate changes, which may lead to a discontinuous update law. To address this issue, this section develops two plant-observer schemes, which can guarantee that the parameter update law is continuous.

Rewrite the plant model (\ref{eq:str-fb-sys}) as
\begin{flalign}\label{eq-parameter-x-model}
    \dot{x}=f(x,u)+F(x)^\T\theta
\end{flalign}
where
\begin{flalign}
    &f(x,u)
    =
    \left[
      \begin{array}{cccc}
        x_2 & \cdots & x_n & u \\
      \end{array}
    \right]^\T
        \\
    &
    F(x)
    =
    \left[
      \begin{array}{cccc}
        \varphi_1(x_1)^\T & \cdots & \varphi_{n-1}(\bar{x}_{n-1})^\T & \varphi_n(x)^\T \\
      \end{array}
    \right]^\T
    \label{eq:def-smallF-and-bigF}
\end{flalign}

\subsection{$x$-Passive Identifier}

Introduce the observer
\begin{flalign}
    \dot{\hat{x}}
    &=\big(A_0-\sigma F(x)^\T F(x)P\big)(\hat{x}-x)
        +f(x,u)+F(x)^\T\hat\theta
        \label{eq:observer-x-model}
\end{flalign}
where $\sigma>0$ is a design parameter, $A_0$ is a matrix defined in (\ref{eq:def-matrix-AC}), and $P$ is the solution to the Lyapunov equation (\ref{eq:Ac-lya-eq}). Then the observer error
\begin{flalign}\label{eq:epsilon-x-passive}
    \epsilon=x-\hat{x}
\end{flalign}
is governed by
\begin{flalign}\label{eq:x-model-obser-err}
    \dot{\epsilon}
    =\big(A_0-\sigma F(x)^\T F(x)P\big)\epsilon+F(x)^\T\tilde\theta.
\end{flalign}
Design the parameter update law
\begin{flalign}\label{eq:x-model-update-law}
    \dot{\hat\theta}
    =\Gamma F(x)P\epsilon, \;\;\;\;\Gamma=\Gamma^\T>0.
\end{flalign}

The following result was given in \cite[Lemma 5.12]{krstic1995nonlinear}.

\begin{lemma}\label{lem:x-passive-bd}
Let $V=\epsilon^T P\epsilon+\tilde\theta^T\Gamma^{-1}\tilde\theta$. Suppose that the solutions of (\ref{eq-parameter-x-model}), (\ref{eq:x-model-obser-err}), and (\ref{eq:x-model-update-law}) are defined on $[0,t_f)$. Then the following properties holds:
\begin{itemize}
\item[(i)] $\tilde{\theta}\in L_\infty[0,t_f)$.
\item[(ii)] $\epsilon\in L_2[0,t_f)\bigcap L_\infty[0,t_f)$.
\item[(iii)] $\dot{\hat{\theta}}\in L_2[0,t_f)$.
  \item[(iv)] $\|\tilde\theta\|_\infty\leq\sqrt{\bar\lambda(\Gamma)V(0)}$ and $\|\dot{\hat\theta}\|_2\leq\frac{\bar\lambda(\Gamma)}{\sqrt{2\sigma}}\sqrt{V(0)}$; moreover, for the case of $\hat h(0)=h(0)$ and $\Gamma=\gamma I$, $\|\tilde\theta\|_\infty\leq|\tilde\theta(0)|$
      and
      $\|\dot{\hat\theta}\|_2\leq\sqrt{\frac{\gamma}{2\sigma}}|\tilde\theta(0)|$.
\end{itemize}
\end{lemma}

Analogous to the schemes above, one can set $\hat{x}(0)=x(0)$ to make the safety violation bound $h_1^*$ smaller. Combining Lemma \ref{lem:ISSf-state-feedback} (ii) and Lemma \ref{lem:x-passive-bd}, we have the following result.

\begin{theorem}[$x$-Passive]\label{thm:x-passive}
For the closed-loop adaptive system consisting of the plant (\ref{eq:str-fb-sys}), controller (\ref{eq:state-fb-controller}), and observer (\ref{eq:observer-x-model}), and the update law (\ref{eq:x-model-update-law}), the following properties hold:
\begin{itemize}
\item[(i)] All signals in the closed-loop system are globally uniformly bounded, and globally asymptotic tracking is achieved.
\item[(ii)]
    For choice of $c_1$, \ldots, $c_n$ satisfying (\ref{eq:choice-of-ci}), if $h_1(0)\geq0$, $\hat x(0)=x(0)$, and $\Gamma=\gamma I$, then $h_1(t)$ satisfies (\ref{eq:zi-L_infty-bd}) with $h_1^*$ given in (\ref{eq:passive-cls-loop-viol-bd}).
\end{itemize}
\end{theorem}

\noindent
\textbf{Proof.} Property (ii) follows by combining Lemma \ref{lem:ISSf-state-feedback} (ii) and Lemma \ref{lem:x-passive-bd} (iv). Now, we prove property (i). By the same argument as in the proof of Theorem \ref{thm:adaptive-state-fb}, all closed-loop signals are bounded. The rest is to prove the convergence of $h_1(t)$. With (\ref{eq:def-W}) and (\ref{eq:def-smallF-and-bigF}),
\begin{flalign}
    W^\T(h,\hat\theta,t)
    &=
    \left[
      \begin{array}{cccc}
        1 & 0 & \cdots & 0 \\
        -\dfrac{\partial\alpha_1}{\partial x_1} & 1 & \ddots & \vdots \\
        \vdots & \ddots & \ddots & 0 \\
        -\dfrac{\partial\alpha_{n-1}}{\partial x_1} & \cdots & -\dfrac{\partial\alpha_{n-1}}{\partial x_{n-1}} & 1 \\
      \end{array}
    \right]F(x)^\T
        \nonumber\\
    &:= N(h,\hat\theta,t)F^\T(x)
\end{flalign}
Let $\zeta=h-N\epsilon$. Then it follows from (\ref{eq:dot-hat-z-state-fb}) and (\ref{eq:x-model-obser-err}) that
\begin{flalign}
    \dot{\zeta}
    =A\zeta
        + [AN-\dot{N}-N(A_0-\sigma F^\T FP)]\epsilon
        + Q^\T\dot{\hat\theta}.
\end{flalign}
For the $i$-th element of $\zeta$, we have
\begin{flalign}\label{eq:diff-zetai-square}
    &\frac{d}{dt}\bigg(\frac{1}{2}\zeta_i^2\bigg)
        \nonumber\\
    &=-s_i\zeta_i^2
        +\zeta_i\zeta_{i+1}
        -\zeta_i\frac{\partial\alpha_{i-1}}{\partial\hat\theta}\dot{\hat\theta}
        \nonumber\\
    &\;\;\;\;+\zeta_i[AN-\dot{N}-N(A_0-\sigma F^\T FP)]_{(i)}\epsilon
        \nonumber\\
    &\leq-\frac{c_i}{4}\zeta_i^2
        -\frac{c_i}{4}\zeta_i^2
        +\zeta_i\zeta_{i+1}
        -g_i\Bigg|\frac{\partial\alpha_{i-1}}{\partial\hat\theta}\zeta_i\Bigg|^2
        +\Bigg|\frac{\partial\alpha_{i-1}}{\partial\hat\theta}\zeta_i\Bigg||\dot{\hat\theta}|
        \nonumber\\
    &\;\;\;\;
        -\frac{c_i}{2}\zeta_i^2
        +|\zeta_i|\Big|[AN-\dot{N}-N(A_0-\sigma F^\T FP)]\Big|_2|\epsilon|
        \nonumber\\
    &\leq-\frac{c_i}{4}\zeta_i^2
        +\frac{1}{c_i}\zeta_{i+1}^2
        +\frac{1}{4g_i}|\dot{\hat\theta}|^2
        \nonumber\\
    &\;\;\;\;
        +\frac{1}{2c_i}\Big|[AN-\dot{N}-N(A_0-\sigma F^\T FP)]\Big|_2^2|\epsilon|^2
\end{flalign}
where $\zeta_{n+1}:=0$. Using the similar argument as in \cite[p. 216]{krstic1995nonlinear}, we can show that $AN-\dot{N}-N(A_0-\sigma F^\T FP)$ is bounded. Also note that $\epsilon,\dot{\hat\theta}\in L_2$. By applying \cite[Lemma B.5]{krstic1995nonlinear} to (\ref{eq:diff-zetai-square}), we have $\zeta_{n}\in L_2$. Similarly, one obtains that $\zeta_1,\ldots,\zeta_{n-1}\in L_{2}$, and hence, $\zeta=[\zeta_1,\ldots,\zeta_n]^\T\in L_2$. Because $N$ is bounded, we get that $h=\zeta+N\epsilon\in L_2$. Then, it follows from Barbalat's lemma that $h(t)$, $\epsilon(t)\rightarrow0$ as $t\rightarrow+\infty$, which establishes the convergence of $h_1(t)$.
\hfill $\Box$
\vskip5pt

\subsection{$x$-Swapping Identifier}

Introduce the following swapping identifier
\begin{flalign}
    \dot{\Omega}^\T
    &=\Big(A_0-\sigma F(x)^\T F(x)P\Big)\Omega^\T+F(x)^\T,\;\;\Omega\in\R^{p\times n}
        \label{eq:x-swap-big-omega}\\
    \dot{\Omega}_0
    &=\Big(A_0-\sigma F(x)^\T F(x)P\Big)(\Omega_0+x)-f(x,u),\;\;\Omega_0\in\R^{n}
        \label{eq:x-swap-big-omega0}
\end{flalign}
where $\sigma>0$ is a design parameter, $A_0$ is a matrix defined in (\ref{eq:def-matrix-AC}), and $P$ is the solution to the Lyapunov equation (\ref{eq:Ac-lya-eq}). Design the parameter update law as
\begin{flalign}
    \dot{\hat\theta}=\Gamma\frac{\Omega\epsilon}{1+\nu|\Omega|_F^2},\;\;\epsilon=x+\Omega_0-\Omega^\T\hat\theta,
    \label{eq:x-swap-gradient-hat-theta}
\end{flalign}
where $\Gamma=\Gamma^\T>0$ and $\nu\geq0$. The following lemma was given in \cite[Lemma 6.5]{krstic1995nonlinear}.

\begin{lemma}\label{lem:bd-adaptive-signal-x-swap}
Let $V=\tilde\epsilon^T P\tilde\epsilon+\tilde\theta^T\Gamma^{-1}\tilde\theta$. Suppose that the solutions of (\ref{eq:x-swap-big-omega}), (\ref{eq:x-swap-big-omega0}) and (\ref{eq:x-swap-gradient-hat-theta}) are defined on $[0,t_f)$.
Then the following properties hold:
\begin{itemize}
\item[(i)] $\tilde{\theta}\in L_\infty[0,t_f)$.
\item[(ii)] $\epsilon\in L_2[0,t_f)\bigcap L_\infty[0,t_f)$.
\item[(iii)] $\dot{\hat\theta}\in L_2[0,t_f)\bigcap L_\infty[0,t_f)$.
\item[(iv)] $\|\tilde\theta\|_\infty\leq\sqrt{\bar\lambda(\Gamma)V(0)}$ and $\|\dot{\hat\theta}\|_\infty\leq\frac{\gamma}{\nu}|\tilde\theta(0)|$. Moreover, for the case of $\Omega(0)=0$, $\Omega_0(0)=-x(0)$ and $\Gamma=\gamma I$,
    $\|\tilde\theta\|_\infty\leq|\tilde\theta(0)|$.
\end{itemize}
\end{lemma}

Set $\Omega(0)=0$ and $\Omega_0(0)=-x(0)$ to eliminate the initial observer error $\epsilon(0)$ on the safety violation bound $h_1^*$.

\begin{theorem}[$x$-Swapping]
For the closed-loop adaptive system consisting of the plant (\ref{eq:str-fb-sys}), controller (\ref{eq:state-fb-controller}), filters (\ref{eq:x-swap-big-omega}), (\ref{eq:x-swap-big-omega0}), and update law (\ref{eq:x-swap-gradient-hat-theta}), the following properties hold:
\begin{itemize}
\item[(i)] All signals in the closed-loop system are globally uniformly bounded, and globally asymptotic tracking is achieved.
\item[(ii)]
For choice of $c_1$, \ldots, $c_n$ satisfying (\ref{eq:choice-of-ci}), if $h_1(0)\geq0$, $\Omega(0)=0$, $\Omega_0(0)=-x(0)$, and $\Gamma=\gamma I$, then $h_1(t)$ satisfies (\ref{eq:zi-L_infty-bd}) with $h_1^*$ given in (\ref{eq:h-swap-h-star}).
\end{itemize}
\end{theorem}

\noindent
\textbf{Proof.}
Property (i) can be established with the proof of \cite[Theorem 6.7]{krstic1995nonlinear}, while property (ii) follows from the combination of Lemma \ref{lem:ISSf-state-feedback} (i) and Lemma \ref{lem:bd-adaptive-signal-x-swap}.
\hfill $\Box$

\subsection{Application to Safety-Critical Tracking under Nonovershooting Constraints}

In the following, we use an example to illustrate the effectiveness of the $x$-passive scheme in Theorem \ref{thm:x-passive}. Let $\Gamma=\gamma I$. Then (\ref{eq:x-model-update-law}) is rewritten as
\begin{flalign}\label{eq:x-model-update-law-xpass-sim}
    \dot{\hat\theta}
    =\gamma F(x)P\epsilon.
\end{flalign}

\begin{example}\label{example-x-passive}\em
Consider system (\ref{eq:example-system}) given in Example \ref{example:z-passive}. The simulation results are given in Figures \ref{Fig:x-passive-time-varying-boundary}-\ref{Fig:xPassive-safety-violation-identifier-para}. Clearly, the $x$-passive parameter estimate $\hat{\theta}$ is smoother than that of the $h$-passive scheme. Comparing Figure \ref{Fig:x-passive-time-varying-boundary} with Figure \ref{Fig:z-passive-time-varying-boundary}, we observe that the adaptation is not frozen for any $t\geq0$. Since the $x$-passive adaptation is activated at system initialization to identify the unknown parameter, the $x$-passive scheme possesses a stronger capability to reduce uncertainty compared to the $h$-passive scheme. Moreover, as shown in Figure \ref{Fig:x-passive-diff-hatTheta-init}, the $x$-passive scheme does not turn off adaptation even when $\hat\theta(t)$ is poorly initialized. Finally, Figure \ref{Fig:xPassive-safety-violation-controller-para} and Figure \ref{Fig:xPassive-safety-violation-identifier-para} demonstrate that, under the $x$-passive scheme, the relationship between the safety violation bound $h_1^*$ and the design parameters is consistent with the one estimated in Theorem \ref{thm:x-passive} (ii). Moreover, we observe that when the parameter estimate $\hat{\theta}(t)$ oscillates around the true value, increasing the adaptation gain $\gamma$ can accelerate the adaptation but also amplifies the oscillation of $\hat{\theta}(t)$, which may lead to a larger safety violation bound.
\begin{figure}[!htb]
 \centering
 \includegraphics[width=6cm]{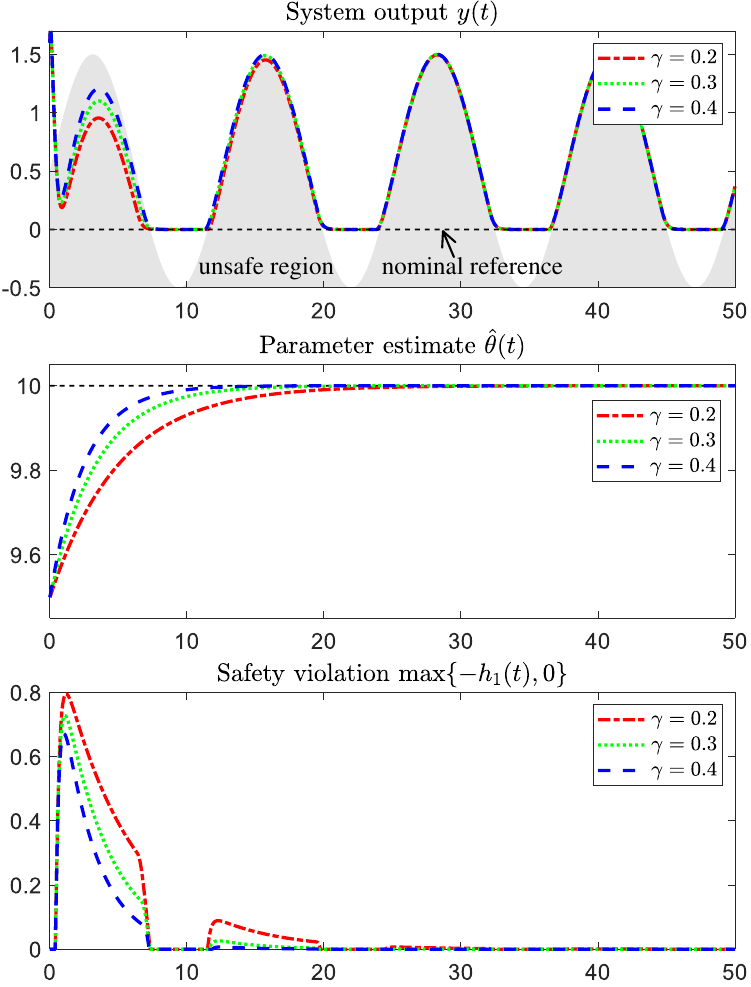}
 \caption{Responses of system (\ref{eq:example-system}) under the $x$-passive scheme with $\sigma=1$, $\hat\theta(0)=9.5$, $y_r(t)\equiv0$, and $r(t)=\sin(t/2) + 0.5$.}
\label{Fig:x-passive-time-varying-boundary}
\end{figure}
\begin{figure}[!htb]
 \centering
 \includegraphics[width=6cm]{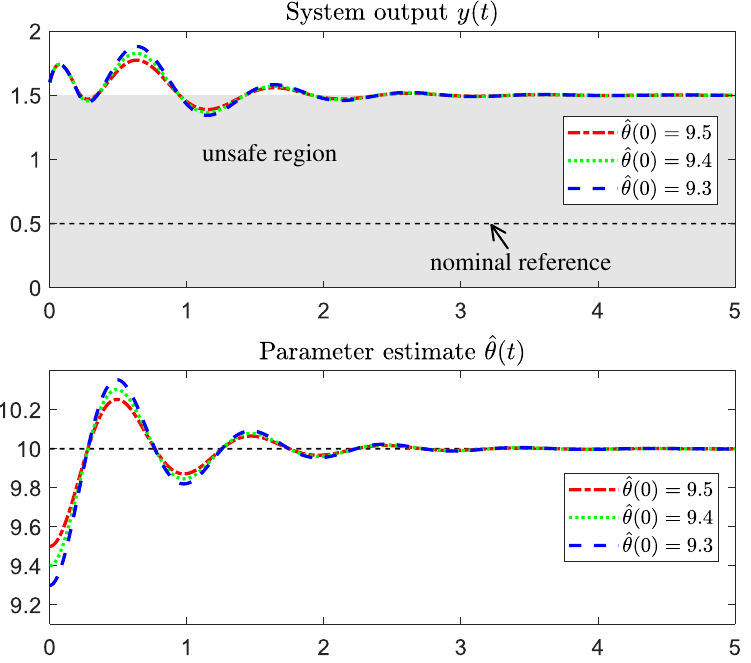}
 \caption{Effect of initial parameter estimate $\hat\theta(0)$ under the $x$-passive scheme with $y_r(t)\equiv0.5$ and $r(t)\equiv1.5$.}
\label{Fig:x-passive-diff-hatTheta-init}
\end{figure}
\begin{figure}[!htb]
 \centering
 \includegraphics[width=9cm]{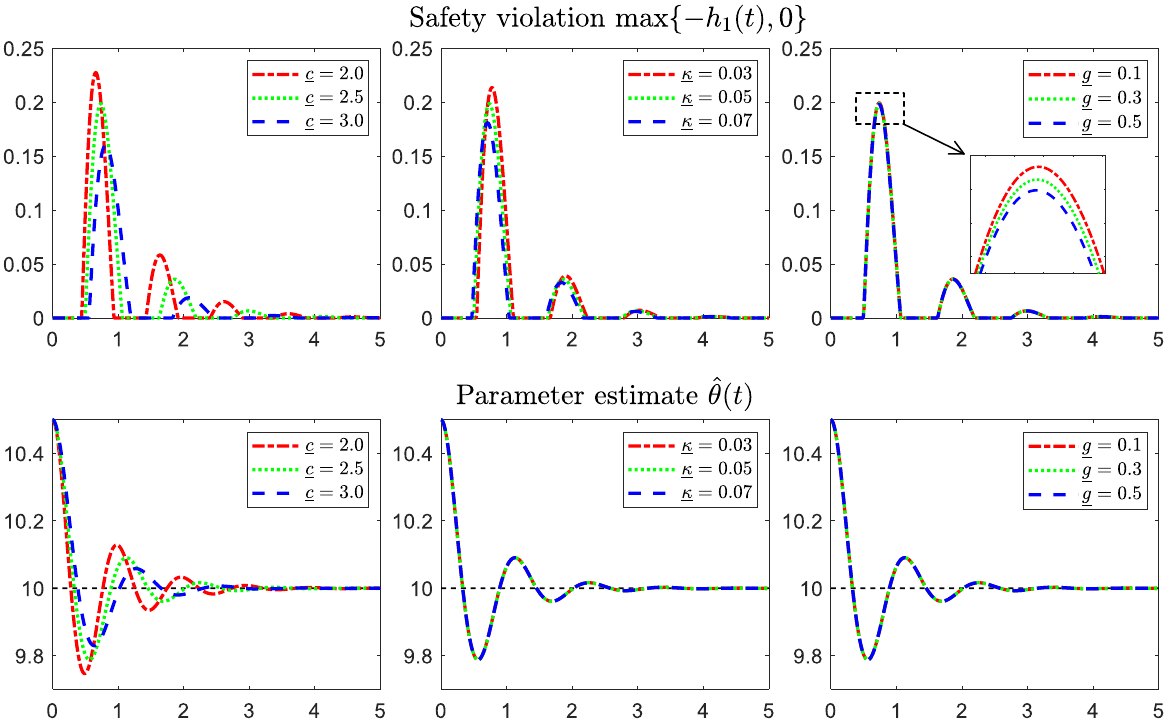}
 \caption{Dependence of safety violation on the controller parameters $\underline{c}$, $\underline{\kappa}$ and $\underline{g}$ in (\ref{eq:underline-c-kappa-g}) under the $h$-passive scheme with $\hat\theta(0)=10.5$, $y_r(t)\equiv0.5$, and $r(t)\equiv1.5$.}
\label{Fig:xPassive-safety-violation-controller-para}
\end{figure}
\begin{figure}[!htb]
 \centering
 \includegraphics[width=6cm]{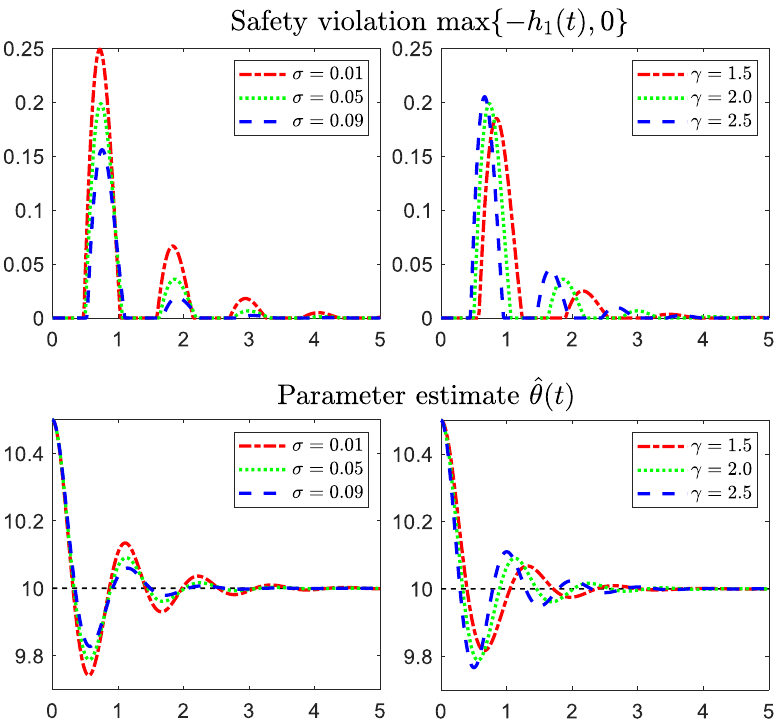}
 \caption{Dependence of safety violation on the identifier parameters $\sigma$ in (\ref{eq:x-model-obser-err}) and $\gamma$ in (\ref{eq:x-model-update-law-xpass-sim}) under the $x$-passive scheme with $\hat\theta(0)=10.5$, $y_r(t)\equiv0.5$, and $r(t)\equiv1.5$.}
\label{Fig:xPassive-safety-violation-identifier-para}
\end{figure}
\end{example}

\section{Conclusion}

A modular adaptive design has been proposed to address the safety-critical control problem under high-relative-degree nonovershooting constraints and parametric uncertainties. The backstepping-based controller module handles the nonovershooting constraint of high relative degree, while the parameter identifier module reduces the parametric uncertainty using online-acquired information about the unknown parameters. This controller-identifier separation offers two advantages. First, it simplifies the controller design procedure by avoiding coupling with high-order derivatives of the parameter estimation error. Second, it provides flexibility in selecting the update law. Because the identifier is separated from the controller module, the update law can be designed with the plant dynamics, rather than being constrained by the gradient of an auxiliary function introduced by the designer (e.g., a CLF or CBF) during controller design. This advantage is particularly important for safety-critical control, where---even in the single CBF case---the overall controller design often involves two or more auxiliary functions (the CLF of the nominal controller is another one). Simulation results demonstrate that the plant-model-based identifier provides a smoother parameter estimate and more effectively reduces uncertainty than its counterpart based on the CBF backstepping error system.


\appendix

\section{Proof of Proposition \ref{prop:delta-epsilon-KL-convergence}}\label{appendix-A}

The convergence of $h_1(t)$ implies that, for any $\varepsilon>0$, there is a $T>0$ such that $|h_1(t)|<\varepsilon$ when $t\geq T$. Let
\begin{flalign}
    A_\varepsilon=\{T\geq0:\forall t\geq T,|h_1(t)|\leq\varepsilon\}.
\end{flalign}
Then, for any $\varepsilon_1\leq\varepsilon_2$, we have $A_{\varepsilon_1}\subseteq A_{\varepsilon_2}$. Define $\bar{T}(\varepsilon)=\inf A_\varepsilon$. By the definition of $A_\varepsilon$ and uniqueness of solutions, we have: i) $\bar{T}(\varepsilon)$ is decreasing (not necessarily strictly); ii) $\lim_{\varepsilon\rightarrow0}\bar{T}(\varepsilon)=+\infty$; and iii) there is a $\bar\varepsilon>0$ such that $\bar{T}(\varepsilon)=0$ for all $\epsilon\geq\bar\varepsilon$ ($\bar\varepsilon$ may be $+\infty$). Similar to the proof of \cite[Lemma 3.1]{lin1996smooth}, take $\hat{T}(\varepsilon)=\bar{T}(\varepsilon)+1/\varepsilon$. Then $\hat{T}(\varepsilon)$ strictly decreases to zero and satisfies $\lim_{\varepsilon\rightarrow0}\hat{T}(\varepsilon)=+\infty$. Hence, by the definition of $\bar{T}(\varepsilon)$,
\begin{flalign}\label{h1-convergence}
    |h_1(t)|<\varepsilon,\;\;\;\;\forall t\geq T(\varepsilon)\geq\bar{T}(\varepsilon).
\end{flalign}
Let $\psi(t)=T^{-1}(t)$. Then $\psi:\R_{>0}\rightarrow\R_{>0}$ is continuous, onto, and strictly decreasing. By taking $\varepsilon=\psi(t)$ and substituting it into (\ref{h1-convergence}), we have
\begin{flalign}\label{h1-convergence-psi}
    |h_1(t)|<\psi(t),\;\;\forall t>0.
\end{flalign}
Recall (\ref{eq:h1-low-bd-in-formulation}) and let
\begin{flalign}
    \bar{\psi}(s,t)=\min\{\rho(s),\psi(t)\}+\frac{a\rho(s)}{1+t}
\end{flalign}
where $a>0$ is an arbitrary constant.
Clearly, $s\mapsto\bar{\psi}(s,t)$ is strictly increasing, $t\mapsto\bar{\psi}(s,t)$ decreases to zero, and $\bar{\psi}(s,t)\leq\bar{\psi}(s,0)=(a+1)\rho(s)$. Hence, there is a $KL$-function $\beta$ satisfying $\beta(s,0)=(a+1)\rho(s)$ and $\beta(s,t)\geq\bar{\psi}(s,t)$. Combining (\ref{eq:h1-low-bd-in-formulation}) and (\ref{h1-convergence-psi}),
\begin{flalign}
    h_1(t)
    &\geq-\bar{\psi}(|\tilde{\theta}(0)|,t)
        \nonumber\\
    &\geq-\beta(|\tilde{\theta}(0)|,t),
    \;\;\;\;\forall h_1(0)\geq0,\;\;\forall t\geq0
\end{flalign}
which completes the proof.

\end{document}